\documentclass[prx,aps,secnumarabic,amsmath,amssymb,twocolumn,notitlepage,longbibliography]{revtex4-2}
\usepackage[normalem]{ulem}
\usepackage[utf8]{inputenc}
\usepackage[english]{babel}
\usepackage[T1]{fontenc}
\usepackage{mathtools}
\usepackage{mathrsfs}
\usepackage{amsfonts}
\usepackage{amsthm}
\usepackage{xcolor}
\usepackage{bbold}
\usepackage{bm}
\usepackage{xr}
\usepackage[unicode=true,breaklinks=false,colorlinks=true,citecolor=black,
linkcolor=black,urlcolor=blue]{hyperref}

\newcommand{\kB}{\kappa_{\rm B}}
\newcommand{\kG}{\kappa_{\rm G}}
\newcommand{\kF}{\kappa_{\rm F}}
\newcommand{\ftn}{f_{\rm d}}
\newcommand{\fnn}{f_{\rm d}^{n}}
\newcommand{\fnm}{f_{\rm e}^{n}}
\newcommand{\gammaeff}{\gamma_{\rm eff}}
\newcommand{\kappaeff}{\kappa_{\rm eff}}

\begin{document}

\title{Theory of defect-mediated morphogenesis}
\author{Ludwig A. Hoffmann$^1$}
\author{Livio Nicola Carenza$^1$}
\author{Julia Eckert$^2$}
\author{Luca Giomi$^1$}
\email{giomi@lorentz.leidenuniv.nl}
\affiliation{$^1$ Instituut-Lorentz, Universiteit Leiden, P.O. Box 9506, 2300 RA Leiden, Netherlands}
\affiliation{$^2$ Physics of Life Processes, Leiden Institute of Physics, Universiteit Leiden,  P.O. Box 9506, 2300 RA Leiden, Netherlands}
\date{\today}

\begin{abstract}
Growing experimental evidence indicates that topological defects could serve as organizing centers in the morphogenesis of tissues.
Here, we provide a quantitative explanation for this phenomenon, rooted in the buckling theory of deformable active polar liquid crystals. 
Using a combination of linear stability analysis and computational fluid dynamics, we demonstrate that active layers, such as confined cell monolayers, are unstable to the formation of protrusions in the presence of disclinations. The instability originates from an interplay between the focusing of the elastic forces, mediated by defects, and the renormalization of the system's surface tension by the active flow. The posttransitional regime is also characterized by several complex morphodynamical processes, such as oscillatory deformations, droplet nucleation, and active turbulence. Our findings offer an explanation of recent observations on tissue morphogenesis and shed light on the dynamics of active surfaces in general.
\end{abstract}

\maketitle

\section{Introduction}

The development of multicellular organisms crucially hinges on internal and external mechanical cues which are transduced by the mechanosensing machinery of cells to give rise to system-wide spatiotemporal rearrangements and, eventually, to the formation of early embryonic features \cite{Conte2012,Nelson2016}. These processes comprise a vast spectrum of active and passive forces, whose regulation relies not only on the molecular repertoire of tissue-forming cells but also on their shape, motility, and local organization. In the past decade, the conceptual framework of active matter \cite{Marchetti2013} --- namely, materials whose building blocks can autonomously move and perform mechanical work --- has enormously contributed to classify the arsenal of physical mechanisms behind force generation and collective migration in both eukaryotes and prokaryotes. However, how these physical mechanisms can be harnessed to achieve biological functionality remains a major open question at the crossroad between developmental biology and mechanics \cite{Metselaar2019, Mietke2019a, Wyatt2020, Al-Izzi2021}. 

One of the most fascinating hypotheses in this respect revolves around the possibility that multicellular systems could take advantage of topological mechanisms to create the conditions, in terms of reproducibility and robustness, for the origin and maintenance of life. In particular, defects have recently been identified by several {\em in vitro} studies as potential candidates for the role of ``topological morphogens'' in various biomimetic systems and model organisms. Keber {\em et al}. \cite{Keber2014}, for instance, demonstrated that protocells, consisting of a single layer of microtubule-kinesin active nematic liquid crystal enclosed in a lipid vesicle and powered by adenosine triphosphate, relieve the mechanical stress, originating from the presence of four topologically required $+1/2$ disclinations, by growing persistent tubular protrusions (Fig. \ref{Fig:fig1}A). More recently, the same mechanism has been invoked in experimental \cite{Livshits2017, Braun2018, Maroudas-Sacks2021,Livshits2021} and theoretical \cite{Metselaar2019,Ruske2021,Vafa2021} studies to explain the growth and regeneration of tentacles in {\em Hydra} (Fig. \ref{Fig:fig1}B). The accumulation of extensile stresses in proximity of integer and semi-integer disclinations has also been proposed by Saw {\em et al}. \cite{Saw2017} as a strategy to achieve cell apoptosis and extrusion in epithelial layers (see also Refs. \cite{Loewe2020, Monfared2021} for recent supporting evidence from simulations) and, more recently, by Guillamat {\em et al}. \cite{Guillamat2020} as a route to the formation of multicellular protrusions in confined myoblasts (Fig. \ref{Fig:fig1}C). Long before the importance of defects for the development of non-planar features in tissues had been recognized, blister-like hemicysts --- also known as ``domes'' --- were already routinely observed in many epithelial cell cultures derived from renal cell lines (see e.g. Refs.~\cite{Wyatt2020,Popowicz1986,Arslan2021}). An example of a dome, obtained from a monolayer of epithelial Madin-Darby Canine Kidney (MDCK) cells, was reproduced here and is shown in Fig. \ref{Fig:fig1}D (see also Movie S1 and S2 and Fig. S1). The central region of this dome, where the curvature is larger than elsewhere,  features a multitude of topological defects (i.e. five- and seven-sided cells) with net positive topological charge: i.e. $\sum_{i\in{\rm dome}}(6-c_{i})>0$, with $c_{i}$ the number of sides of the $i$th cell. That is, the system exhibits an excess of positive disclinations (i.e. five-sided cells) where its curvature is maximal. This observation (see Secs. S1 and S2 for further experimental evidence) suggests the possibility that, similar to the other examples illustrated above, the formation of cellular domes could be facilitated by topological defects in combination with other system-specific mechanism, such as the injection of fluid under the cell layer which results in a focal detachment \cite{Cereijido1978, Latorre2018}.

Whereas these experimental studies have now convincingly shown the existence of a correlation between topological defects and certain morphogenic events, a clear theoretical picture of the mechanical nature of this behavior is still missing. Here, we address this lack.
Using a combination of classical differential geometry, linear stability analysis and computational fluid dynamics, we demonstrate that confined active layers, here modeled as active polar liquid crystals, are unstable to the formation of protrusions in the presence of a $+1$ disclination. In extensile systems with positive flow alignment, the instability originates from an interplay between the focusing of elastic forces, mediated by the defect, and a renormalization of the system's surface tension by the active flow, arising in response to the perpetual injection of active stress under confinement. In the posttransitional phase, such competition leads to additional dynamical regimes, which includes oscillatory deformations, the nucleation of droplets, and a turbulent state with proliferating protrusions. By contrast, in contractile systems, the same phenomenon stabilizes the flat configuration, thus preventing the formation of protrusions. 

A precise account of all the examples of defect-mediated morphogenesis illustrated in the introduction requires, in principle, a great deal of biophysical details and a case-by-case approach. However, their very diversity suggests the existence of a general underlying mechanism. Our goal is to identify and rationalize this mechanism in the simplest possible setting, thus making it amenable for an in-depth analytical description.

\section{Results}

\subsection{\label{sec:model}The model}

Our model layer consists of a thin film of active polar or nematic liquid crystal, whose mid-surface $\mathcal{M}$ is spanned by the unit normal $\bm{n}$ and the pair of orthogonal tangent vectors $\bm{g}_{i}$, with $i=1,\,2$, and $g_{ij}=\bm{g}_{i}\cdot\bm{g}_{j}$ the associated metric tensor. The mean and Gaussian curvatures are denoted by $H$ and $K$ respectively (see Fig.~\ref{fig:fig2}A and, e.g., Ref.~\cite{Deserno2015}). The local average orientation of the cells is described by the unit vector $\bm{p}=p^{i}\bm{g}_{i}$, with $|\bm{p}|^{2}=p^{i}p_{i}=1$. The system free energy is given by
\begin{equation}
F = \int_\mathcal{M} {\rm d}A\,\left[ \gamma + \kB H^2 + \kG K + \dfrac{\kF}{2}\,(\nabla_{i}p_{j})(\nabla^{i}p^{j})\right].
\label{eq:FreeEnergy}
\end{equation}
The first three terms on the right-hand side of Eq. ~(\ref{eq:FreeEnergy}) account for the energetic cost of deformations of the mid-surface, where $\gamma$ is the surface tension, $\kB$ is the bending rigidity and $\kG$ is the Gaussian splay modulus~\cite{Helfrich1973}. The last term, on the other hand, is the Frank free energy~\cite{Chaikin1995} expressing the compliance of the system to a local distortion of the cellular polarization, with $\nabla_i$ denoting covariant differentiation and $\kF$ denoting the rotational stiffness in the one-elastic-constant approximation.

\begin{figure}[t]
\centering 
\includegraphics[width=\columnwidth]{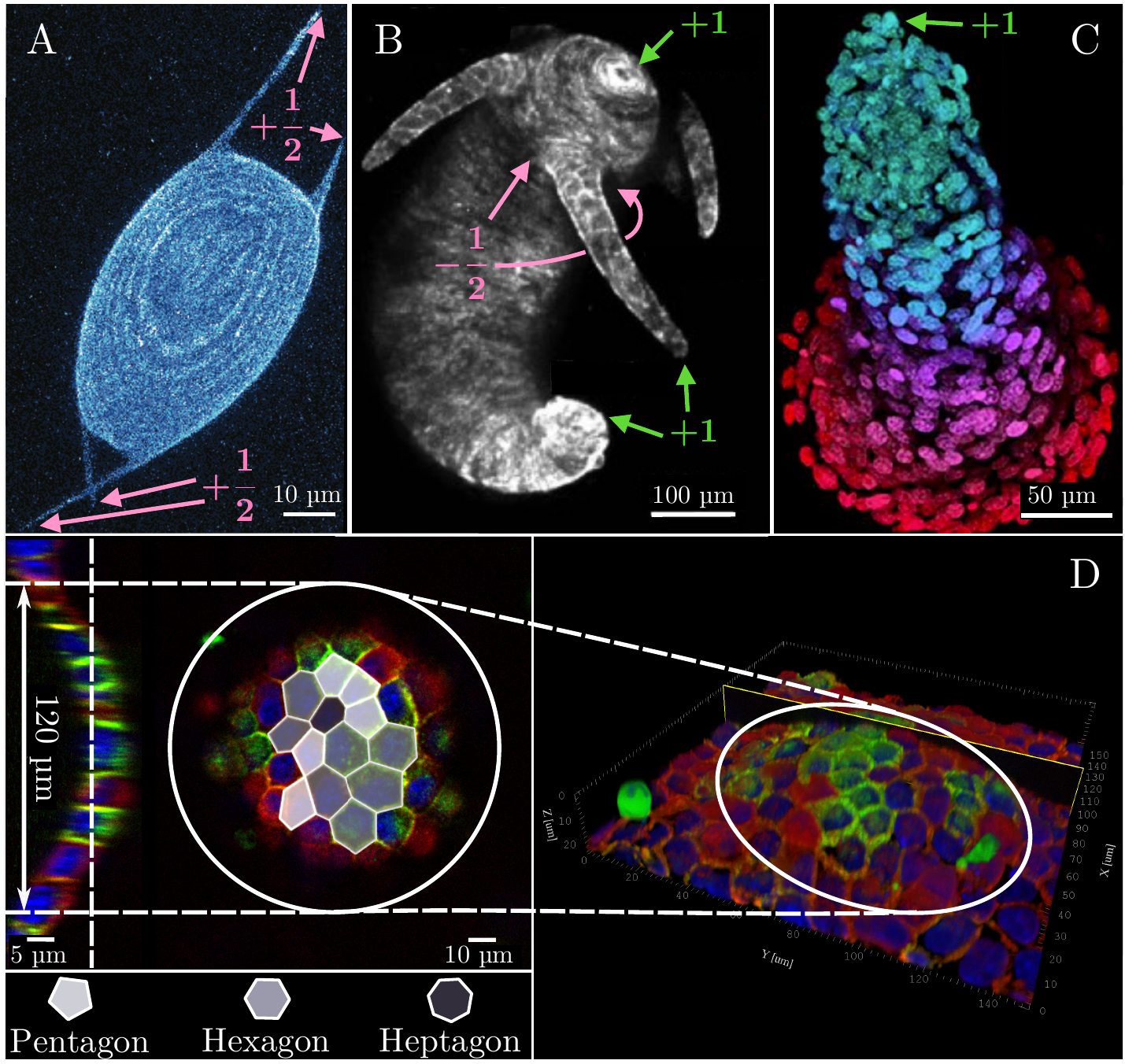}
\caption{\textbf{Examples of morphogenic events in the presence of topological defects.} (A) Active nematic protocell consisting of a monolayer of microtubules and kinesin enclosed in a lipid vesicle. The four topologically required $+1/2$ defects deform the lipid envelope, thereby creating tubular protrusions (adapted from Ref. \cite{Keber2014}). (B) Example of {\em Hydra} featuring $+1$ disclinations in proximity of the mouth, the foot and the tip of each tentacle, and two $-1/2$ defects at the base of each tentacle (adapted from Ref. \cite{Maroudas-Sacks2021}). (C) Multicellular protrusion in layers of collectively migrating myoblasts, featuring a $+1$ disclination at the tip (adapted from Ref. \cite{Guillamat2020}). (D) Example of a dome obtained from an initially flat monolayer of cultured MDCK GII cells. The central region of the dome, where the curvature is larger than elsewhere, hosts several topological defects (i.e. five and seven-sided cells), with a net positive topological charge. On the left-hand side a cross section and a top view of the dome with a superimposed tessellation are shown (red, F-actin; green, E-cadherin; blue, nuclei).}
\label{Fig:fig1}
\end{figure}

The steady state that we aim at describing comprises a dense population of cells freely translating and rotating along a stationary mid-surface $\mathcal{M}$. The former requirements amount to the following set of hydrodynamic equations for $\bm{p}$ and the momentum density $\rho\bm{v}=\rho v^{i}\bm{g}_{i}$ on the tangent plane~\cite{Giomi2015, Pearce2019, Mietke2019b, Santiago2018}:
\begin{subequations}\label{eq:hydrodynamics}
\begin{gather}
\rho(\partial_t + v^{k}\nabla_{k}) v^{i}= \nabla_{j} \left(\sigma_{\rm h}^{ij}+\sigma_{\rm a}^{ij} \right)+\ftn^{i}\;,\\ 
\hspace{-1ex}(\partial_t + v^{k}\nabla_{k}) p^{i}  = \left(g^{ij}-p^ip^j\right)\left(\lambda u_{jk}p^k-\omega_{jk}p^k+\Gamma h_{j}\right).\hspace{-1ex}
\end{gather}
\end{subequations}
The total cellular mass $M=\int {\rm d}A\,\rho$ is assumed to be constant in time, although the number of cells could in principle fluctuate as a consequence of cell division and apoptosis. Thus $\rho={\rm const}$ and $\nabla_i v^i=0$. This assumption is justified by the fact that, in cellular systems, the speed $v$ of migrating cells is orders of magnitude lower than the speed of sound $c_{\rm s}$. Because the typical magnitude of density fluctuations is $\delta\rho\sim{\rm Ma}^{2}$, with ${\rm Ma}=v/c_{\rm s}$ the Mach number, this implies, to first approximation, $\delta\rho \approx 0$. The three terms on the right-hand side of Eq.~(\ref{eq:hydrodynamics}a) correspond, respectively, to the hydrodynamic stress tensor  $\sigma_{\rm h}^{ij} = -P_\text{h} g^{ij} + 2 \eta u^{ij}$ ---where $P_\text{h}$ is the pressure, $\eta$ is the shear viscosity and $u^{ij}=(\nabla^i v^j + \nabla^j v^i)/2$ is the strain-rate tensor--- the active stress $\sigma_{\rm a}^{ij}$ and the force per unit length $\ftn^{i}$, originating from the distortion induced by defects and Gaussian curvature and driving the so-called elastic ``backflow''~\cite{Toth2002}. The latter can be derived from expressing the polarization gradients in terms of the geometric potential $\chi$, subject to the Poisson equation $\nabla^2 \chi = K - \rho_{\rm d}$, where $\rho_{\rm d}$ is the topological charge density~\cite{Bowick2000, Giomi2007, Bowick2009, Santiago2018}. Explicitly, $\ftn^{i} =- 2 \kF\rho_{\rm d}\nabla^{i}\chi$ (see SI). Thus, at equilibrium, $\rho_{\rm d} \sim K$  so that defects  place themselves in regions of like-sign Gaussian curvature to minimize the elastic free energy \cite{Bowick2009}. 
Last, the active stress tensor embodies the contribution of the force dipoles autonomously generated by the cells and, following a standard construction, can be expressed as $\sigma_{\rm a}^{ij} = \alpha (p^{i}p^{j}-g^{ij}/2)$~\cite{Pedley1992, Simha2002,Asano2009,Marchetti2013,You2018}. The phenomenological constant $\alpha$ quantifies the magnitude of the cellular forces and is positive (negative) for contractile (extensile) systems.

Equqation (\ref{eq:hydrodynamics}b) governs the rotational dynamics of the cells which, in turn, is dictated by the coupling with the local flow field, expressed by the first two terms on the right-hand side of the equation, with $\lambda$ the flow alignment parameter and $\omega^{ij}=(\nabla^i v^j - \nabla^j v^i)/2$ the vorticity tensor. The last term is the molecular field $h_{i}=-\delta F/\delta p^{i}=\kF\nabla^{2}p_{i}$ which drives the system toward the ground state of the free energy, with $\Gamma^{-1}$ the rotational viscosity (see e.g. Ref.~\cite{Chaikin1995}).

Last, the stationarity of the mid-surface $\mathcal{M}$ requires the net force acting along the normal direction $\bm{n}$ to vanish, hence
\begin{equation}
K_{ij} \left(\sigma_{{\rm h}}^{ij}+\sigma_{\rm a}^{ij}\right) + \fnn + \fnm = 0\;,
\label{eq:NormalForceBalance}
\end{equation}
where $K_{ij}$ is the curvature tensor (see Fig.~\ref{fig:fig2}A). The normal forces per unit length $\fnn$ and $\fnm$ on the left-hand side of Eq. (\ref{eq:NormalForceBalance}) are found from minimizing the free energy $F$ (see SI) and are given by $\fnm = 2 \gamma H + 2 \kB H (K - H^2) - \kB \nabla^{2} H$ and $\fnn = 2 \kF (2Hg^{ij}-K^{ij})\nabla_i\nabla_j\chi + 2 \kF (K^{ij}-Hg^{ij})\nabla_i\chi\nabla_j\chi$~\cite{Capovilla2002c, Capovilla2004, Guven2006, Salbreux2017, Santiago2018}. We stress that Eq.~(\ref{eq:NormalForceBalance}) reduces to the Helfrich shape equation, $\fnm = 0$, in the limit $\alpha = \kF = 0$~\cite{Capovilla2002c, Deserno2015}, and to the van K\'arm\'an equation, $\fnn + 2 \gamma H = 0$, in the limit $\alpha = \kB = 0$~\cite{Landau1970, Seung1988}.

\subsection{\label{sec:Analysis}Stability of defective active layers: Topology, geometry and morphogenesis}

In this section, we investigate how topological defects and activity conspire to render an initially flat layer unstable to the growth of protrusions. While Eqs. (\ref{eq:hydrodynamics}) and (\ref{eq:NormalForceBalance}) hold for arbitrary conformations of the midsurface $\mathcal{M}$, here we consider the simpler case of an axisymmetric surface and a liquid crystal with polar symmetry, where the local polarization and velocity fields depend solely on the distance $r \le R$ from the symmetry axis, with $R$ the radius of the layer: i.e., $\bm{p}=\bm{p}(r)$ and $\bm{v}=\bm{v}(r)$. Under this assumption, the polarization field inevitably features a $+1$ topological defect at the center of the system. Thus, taking $\rho_{\rm d}=2\pi \delta(\bm{r})$, with $\delta(\cdot)$ a delta function on $\mathcal{M}$, and setting $\chi=0$ at $r=R$ without loss of generality, the geometric potential can readily be found to be  $\chi=-\log(r/R)$.

To make progress, we parameterize the midsurface $\mathcal{M}$ in a Monge patch, so that the position $\bm{r}=\bm{r}(r,\varphi)$ of a generic point is given by $\bm{r}=r\bm{g}_{r}+h\bm{e}_{z}$, with $\bm{e}_{z}$ a unit vector in the $z$ direction and $h=h(r)$ the height. Following a standard approach of membrane physics (see, e.g., Ref. \cite{Deserno2015}), we then assume that $|\nabla h| \ll 1$, so that one can ignore terms of order $\mathcal{O}\left(|\nabla h|^2\right)$ and higher in Eqs. (\ref{eq:hydrodynamics}) and (\ref{eq:NormalForceBalance}). Under such a small-gradient approximation, $g_{ij} \approx \delta_{ij}$, $H\approx-\nabla^{2}h/2$, $K\approx 0$, and Eqs. (\ref{eq:hydrodynamics}) reduce to their flat counterpart. Moreover, axial symmetry and incompressibility yield $v^{r}=0$, so that $\bm{v}=1/r\,v^{\varphi}\bm{g}_{\varphi}$, with $v^{\varphi}=v^{\varphi}(r)$, and $\bm{p} = \cos\epsilon\,\bm{g}_{r}+1/r\,\sin\epsilon\,\bm{g}_{\varphi}$ with $\epsilon$ a constant and $(r,\varphi)$ polar coordinates on the $h=0$ plane. Solving Eq. (\ref{eq:hydrodynamics}b), we find $\epsilon = \pm \arccos(-1/\lambda)/2$ under the assumption that $|\lambda|>1$. Similarly, neglecting unimportant inertial terms~\cite{carenza2020_bif} and solving Eq. (\ref{eq:hydrodynamics}a) for no-slip boundary conditions gives, after standard algebraic manipulations (see e.g. Ref. \cite{Kruse2004, Giomi2014, Hoffmann2020}):
\begin{subequations}\label{eq:sol}
\begin{gather}
v^{\varphi} = \mp \frac{\alpha}{\eta}\,\frac{\sqrt{1-1/\lambda^{2}}}{2}\,r\log\frac{r}{R}\;,\\
P_\text{h} = -\frac{\alpha}{\lambda} \log \frac{r}{R} - \kF \rho_{\rm d}\;,
\end{gather}
\end{subequations}
consistent with a classic result by Kruse {\em et al}.~\cite{Kruse2004}. It is worth stressing that, by virtue of Eq. (\ref{eq:sol}b), the pressure nominally diverges at the origin, where the $+1$ defect is located. Similar to other divergences stemming from the continuous description of defects, however, this one too can be regularized by introducing the coherence length $\ell_{\rm c}$, setting the short length scale cut-off at which the continuous theory breaks down, because the magnitude of the polarization field, here assumed to be constant, drops in proximity of a defect. Thus $\ell_{\rm c} \le r \le R$. Last, replacing Eqs. (\ref{eq:sol}) in Eq. (\ref{eq:NormalForceBalance}), gives a stability condition for the flat conformation of the midsurface $\mathcal{M}$ in the form of a fourth-order linear partial differential equation with position-dependent coefficients:
\begin{equation}
\label{eq:HeightEquation}
\frac{\kB}{2}\,\nabla^{4}h-\gammaeff\nabla^{2}h+\frac{\kappaeff}{r}\,\partial_{r}\left(\frac{\partial_{r}h}{r}\right)=0\;,
\end{equation}
where $\gammaeff=\gamma-P_\text{h}$ and $\kappaeff=\kF+\alpha r^{2}/(2\lambda)$ are the effective surface tension and Frank's elastic constant, respectively, which include the renormalization resulting from the active flow. Note that the velocity field does not enter Eq. (\ref{eq:HeightEquation}) explicitly, as the term $K_{ij}u^{ij}$ vanishes identically. Moreover, $\mathcal{M}$ is clamped at the boundary such that $h$ and all its derivatives vanish at $r=R$. We stress that the assumption of rotational symmetry restricts us to the low-activity regime. Upon increasing the activity, the system would eventually enter a turbulent regime, thereby losing its spatial symmetry. Evidently, this regime cannot be captured by our solution.

\begin{figure}[t!]
\centering
{\includegraphics[width=1.\columnwidth]{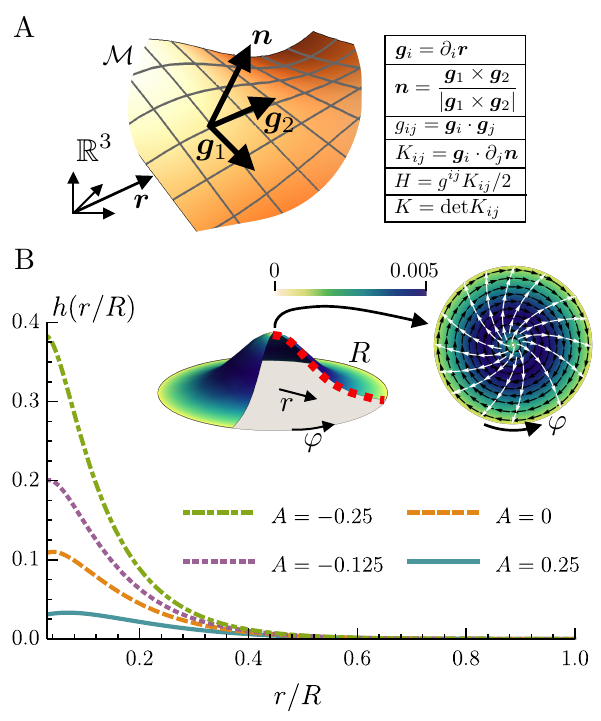}}
\caption{\textbf{Post-buckling geometry of layer.} (A) Sketch of the midsurface $\mathcal{M}$ embedded in $\mathbb{R}^3$. In the table, the cross and the dot refer to the vector and scalar product, respectively, with respect to the Euclidean metric of $\mathbb{R}^3$. In the figure and throughout the article bold letters indicate $\mathbb{R}^3$ vectors and Latin indices indicate surface coordinates on $\mathcal{M}$. The coordinate system on $\mathcal{M}$ is denoted by $\{\bm{g}_i, \bm{n}\}$ and it defines the metric $g_{ij}$ and the curvature tensor $K_{ij}$ of the surface, hence the mean curvature $H$ and Gaussian curvature $K$. (B) In the top middle a sketch of the buckled surface with a $+1$ defect at the center. The disk on the top right is a view from the top. The black arrows are the azimuthal velocity field, the color scale is its magnitude, and the white arrows are the director field. In addition, we solved Eqs. (\ref{eq:hydrodynamics}) and (\ref{eq:NormalForceBalance}) numerically to plot the height profile $h(r/R)$ of the surface for different values of the active number $A = \alpha \ell_{\rm c}^2/\kappa_F$. Comparing the active with the passive ($A = 0$) case, we see how an extensile activity ($A < 0$) favors the buckling, while a contractile one activity inhibits it. The values of the constants used are the following: $\gamma = 0.012$, $\Gamma = 1$, $\lambda = 1.1$, $\kappa_F = 0.02$, $\kappa_B = 0.01$, and $\eta = 5/3$.}
\label{fig:fig2}
\end{figure}

\begin{figure*}[t]
\centering
{\includegraphics[width=1.\textwidth]{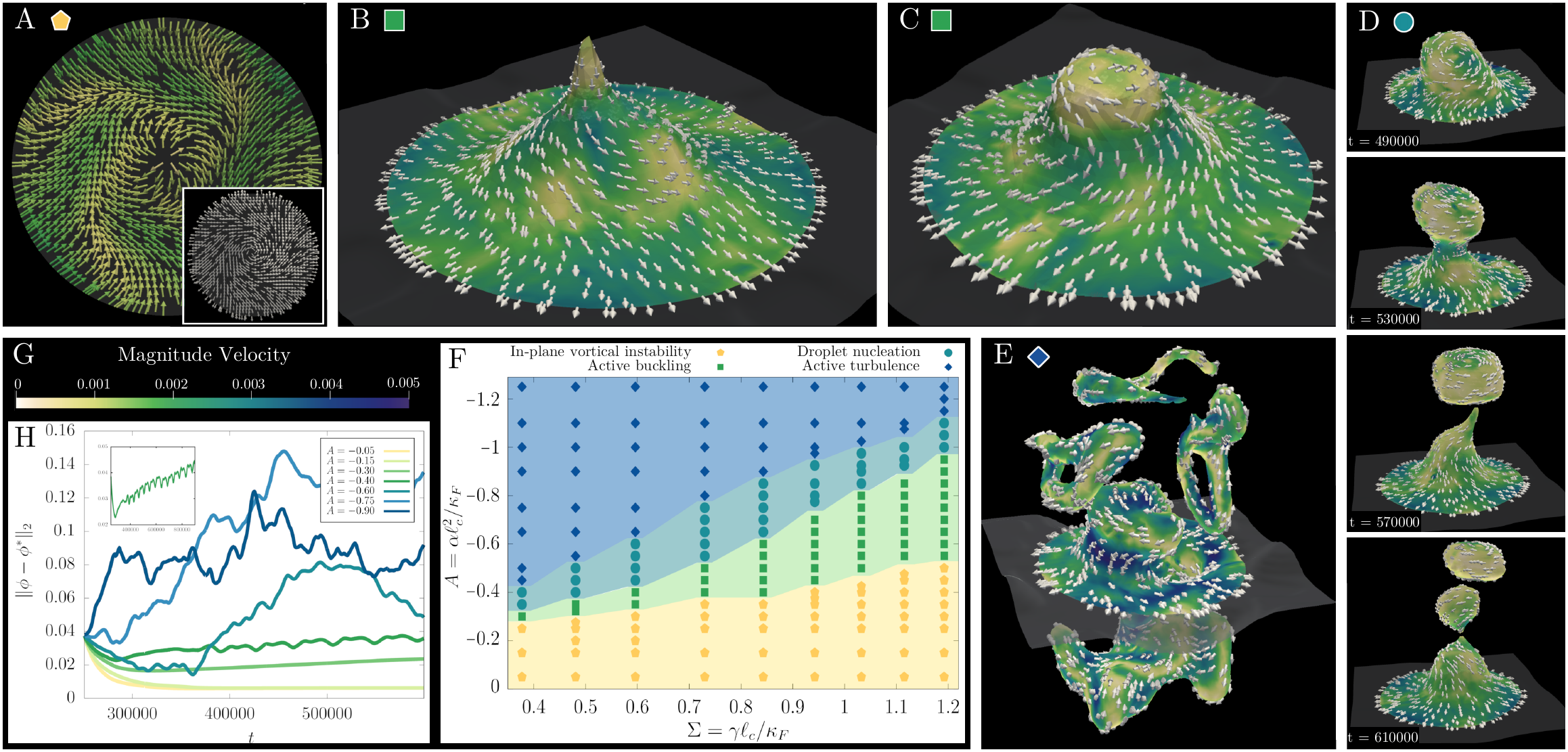}}
\caption{\textbf{Dynamical regimes in active layers.} (A to E) Defect-mediated buckling and activity-driven shape deformations of an active polar layer obtained from numerical simulations (see Sec. \ref{sec:dynamical}). The corresponding phase diagram is shown in (F). In all simulations we fix the dimensionless surface tension $\Sigma=0.6$ and vary the dimensionless activity $A$. The white vectors mark the polarization field, while the color code in (G) refers to the local magnitude of the flow.
At small activity (yellow pentagon in the phase diagram), we find a vortical flow structure and the interface is flat outside the defect core. This is illustrated in (A) where $A=-0.2$ and the inset shows the spiral pattern in the polarization field. Beyond the threshold of the buckling instability (green squares), the layer attains a funnel-like profile with the topological defect at the end of the hollow; this is shown in (B). At slightly higher activity, just below the transition to the region of droplet production ($A=-0.4$), we find that the interface begins to oscillates between this funnel-like state and the configuration shown in (C). Next, in (D) we show snapshots taken at different times during the process of droplet nucleation at $A=-0.55$ (cyan circles). Last, in (E), we show a chaotic configuration of the layer in the active turbulent regime ($A=-1.0$, blue diamonds). (H) The time evolution of the normalized distance in the $L_2$-space between the simulated profile of the phase field $\phi$ and the equilibrium one, $\phi^*=\tanh{(z/\xi)}$, with $\xi$ the equilibrium interface width between the two isotropic fluids and $z$ the coordinate in the direction normal to the interface. The inset on the top left in (H) refers to the case at $A=-0.4$ and shows the periodic oscillations between the configurations of (B) and (C) for $A=-0.4$ in terms of the normalized $L_2$-distance.}
\label{fig:fig3}
\end{figure*}

Now, to gain insight into the stability of the system with respect to the formation of protrusions, we separately consider three limit cases, namely, {\em (i)} $\kB=0$ and $\alpha=0$; {\em (ii)} $\kB=0$ and $\alpha \ne 0$, and {\em (iii)} $\kB\ne 0$ and $\alpha\ne 0$. From a physical perspective, the first two cases correspond to layers of passive and active subunits, respectively, whose elastic stiffness is sufficiently small to be considered negligible, but where surface tension penalizes an increase in the area. The vanishing bending modulus, in particular, renders highly curved features, such as kinks or cusps, energetically acceptable. The first scenario has been considered by Frank and Kardar~\cite{Frank2008} and corresponds to the case of a passive liquid-crystalline disk-like interface plagued by a $+1$ disclination in the center.
In this case Eq. (\ref{eq:HeightEquation}) can be integrated exactly, to give
\begin{equation}\label{eq:reduced_height}
\left(r^{2}-R_{\rm c}^{2}\right) \partial_{r} h = 0\;,
\end{equation}
with $R_{\rm c}^{2}=\kF/\gamma$. Because of our boundary conditions for $h$, this equation admits a nontrivial solution only if $R> R_{\rm c} = \sqrt{\kF/\gamma}$. Even Eq.~(\ref{eq:NormalForceBalance}) is exactly integrable in this case and $\mathcal{M}$ is found to be a parabolic pseudosphere ---namely a surface of constant negative Gaussian curvature--- whose height and mean curvature diverge as $r\rightarrow 0$ \cite{Frank2008}. Similarly, in the second scenario, Eq.~(\ref{eq:HeightEquation}) reduces to Eq.~(\ref{eq:reduced_height}), but with $R_{\rm c}$ given by the solution of the transcendental equation $\kF/R_{\rm c}^{2} =  \gamma - \alpha/(2\lambda) - \alpha/\lambda\log R/R_{\rm c}$. As in the previous case, we find that there is a non-trivial solution only if $R > R_{\rm c}$, thus the flat conformation becomes unstable to the growth of protrusions when the radius $R$ of the active layer exceeds the critical radius
\begin{equation}
R_{\rm c} = \sqrt{\frac{\kF}{\gamma - \frac{\alpha}{2 \lambda}}} \;.
\label{eq:Instability}
\end{equation}
Equivalently, for fixed $R$ values, Eq. \eqref{eq:Instability} provides a threshold in $|\alpha|$, above which the layer becomes unstable to buckling. Note that for $\lambda > 1$ the activity reduces (increases) the effective surface tension in the presence of contractile (extensile) stresses and vice versa for negative $\lambda$ (see also Ref. \cite{Salbreux2017}). Conversely, if $\gamma\le\alpha/(2\lambda)$, then Eq. (\ref{eq:HeightEquation}) has no nontrivial solutions and the flat solution is always stable. 

Last, in the third and most generic scenario, Eq. (\ref{eq:HeightEquation}) cannot be reduced to Eq. (\ref{eq:reduced_height}) and one cannot find a stability criterion as simple as that expressed by Eq. (\ref{eq:Instability}). However, as for the passive instability discussed in Ref. \cite{Frank2008}, we expect the bending stiffness $\kB$ to merely renormalize Frank's elastic constant $\kF$. In addition, a finite bending rigidity makes cusps, kinks and other singularities characterized by a diverging curvature energetically prohibitive; hence we expect the tip of the protrusion to become increasingly smooth with increasing $\kB$. To substantiate the latter statement, we have lifted the small-gradient approximation and numerically integrated Eqs.~(\ref{eq:hydrodynamics}) and (\ref{eq:NormalForceBalance}) on an axisymmetric surface. The equations do not decouple in this limit, therefore it is not possible to condense them in a single equation for the height $h$. However, one finds from the incompressibility equation and with $\bm{p}=1/g\,\cos\epsilon\,\bm{g}_{r}+1/r\,\sin\epsilon\,\bm{g}_{\varphi}$, where $g=\sqrt{1+(\partial_{r}h)^{2}}$, that $\bm{v}=1/r\,v^{\varphi}\bm{g}_{\varphi}$, and the angle $\epsilon$ as well as the pressure $P_\text{h}$ are left unchanged. The remaining fields, $v^{\varphi}$ and $h$, can be numerically computed and $h$ is shown in Fig. \ref{fig:fig2}B for different values of activity. Note that the solutions found for $v^{\varphi}$ are essentially identical to the analytical solutions obtained using the small-gradient expansion.

In summary, the presence of a $+1$ defect renders the planar configuration of an active polar layer unstable to buckling. The instability arises from the fact that the defect introduces an angular deficit in the orientation of the polarization field that a like-sign Gaussian curvature can compensate, thereby mitigating the distortion sourced by the defect \cite{Bowick2009}. Despite having a passive origin, this instability is affected by the hydrodynamic flow fueled by the layer's extensile or contractile activity. This flow changes the later pressure acting on an arbitrary fluid patch, resulting in an additional compression (for extensile activity) or stretching (for contractile activity) of the layer, which, in turn, inevitably interferes with how the layer itself deforms out of plane, thus making the system respectively more or less probe to buckling.

\subsection{\label{sec:dynamical}Buckling, oscillatory regime and droplet nucleation}

The interplay between defect-mediated stress focusing and the elasticity of the active layer, illustrated in the previous section, is a notable example of how the interplay between the topology of the polarization field and the geometry of the flow cooperatively render an initially flat layer unstable to the growth of protrusions. In this section, we look at the evolution of the instability to unveil how topological defects influence the morphology of the protrusion in the post-buckling scenario. To this end, we numerically simulate an active polar layer, whose thickness $\xi$ is comparable in magnitude with the coherence length $\ell_{\rm c}$, which, in turn, represents the shortest length scale in our continuum description. The active layer is sandwiched between two Newtonian fluids, whose relative concentration is described by the phase field $\phi=\phi(\bm{r})$~\cite{Carenza22065,carenza2020_physA}. In the limit $\xi \to 0$, the simulated interface can be factually interpreted as the mid-surface $\mathcal{M}$ of Sec.~\ref{sec:model}~\cite{Liu2003,Ruske2021}. A description of the model is provided in the Materials and Methods section and SI.  We numerically integrate the three-dimensional (3D) hydrodynamic equations of this diffuse interface model in a cylindrical container with homeotropic boundary conditions along the base, so that the equilibrium configuration in the absence of activity is stationary and characterized by a $+1$ disclination at the center of the disk. The interface is slightly deformed at the defect position to accommodate the tendency of the liquid crystal to escape in the normal direction. Outside the defect core $\ell_c$ the interface is flat, in agreement with our analytical prediction in Sec. \ref{sec:Analysis}, since, in the absence of activity, we have buckling only in a disk of radius $R_{c} = \sqrt{5/3}\,\ell_c$. In the following we will restrict our analysis to extensile systems as we find that contractile ones do inhibit a buckling instability, in agreement with what is predicted by our analytical argument. The main control parameters that we varied in our numerical experiments are \emph{(i)} the dimensionless activity $A=\alpha \ell_{\rm c}^2/\kF$ and \emph{(ii)} the reduced surface tension $ \Sigma = \gamma \ell_{\rm c} /\kF$. 

As the dimensionless activity $A$ is increased, four different regimes are encountered. At small $A$ values, the surface is flat, whereas the polarization evolves into a spiral configuration coupled to a vortex flow (Fig.~\ref{fig:fig3}A), in agreement with our prediction in Sec.~\ref{sec:Analysis} and Ref.~\cite{Kruse2004}. As activity is further increased, the interface undergoes a buckling instability (light green squares in Fig.~\ref{fig:fig3}F), which results in the development of a protrusion, featuring a $+1$ defect at the tip. Thanks to our computational approach, we notice that once the instability is set by the previously described competition between defect-mediated stress focusing and elasticity, the development of nonplanar features is further facilitated by the lengthening of the vortical flow field out of plane, a process called \emph{vortex stretching}, which occurs in 3D fluids as the result of conservation of angular momentum~\cite{drazin_1992}. The transition line between the flat and buckled configurations in the phase diagram of Fig.~\ref{fig:fig3}F shows a linear behavior in the $A-\Sigma$ plane, consistent with the analytical criterion expressed by Eq. (\ref{eq:Instability}). 

The competition between active and elastic stresses may eventually give rise to large perturbations around the stationary cuspidal profile, with the surface oscillating from the singular configuration with negative curvature of Fig.~\ref{fig:fig3}B to the smooth configuration of Fig.~\ref{fig:fig3}C. The dynamic of the oscillations can be captured by measuring the temporal evolution of the distance of the simulated interface from the flat configuration, shown in Fig.~\ref{fig:fig3}H, and its inset for the case at $A = - 0.4$ and $\Sigma=0.6$ (see also Movie S3). The frequency of the oscillation linearly increases with $|\alpha|$ along the transition line, which can be rationalized either by dimensional analysis or from the balance of non-equilibrium stresses and viscous ones. This oscillatory instability lies at the heart of another notable behavior which is observed when activity is further increased. In this case, the surface's entropic elasticity is no longer able to counteract the active stresses fueling the growth of the protrusion (top panels in Fig.~\ref{fig:fig3}D and Movie S4). This results in a steady increase of the passive stresses along the protrusion, which eventually results in the breaking of the interface and the consequent nucleation of a droplet (bottom panels in Fig.~\ref{fig:fig3}D and Movie S4). The emulsified droplets are therefore enclosed in a thin active polar shell, which separates the interior of the droplet from the outer Newtonian fluid. For each droplet, the polarization field develops two boojums (i.e. $+1$ defects) as required by the Gauss-Bonnet theorem. 

In the limit of very large activity (blue diamonds in Fig.~\ref{fig:fig3}F), the dynamics becomes fully chaotic, a regime which is known as \emph{active turbulence} in the literature~\cite{Giomi2015,Alert2020,carenza2020,carenza2020_bif}. The active layer is characterized by the proliferation of many amorphous protrusions (see Fig.~\ref{fig:fig3}E and Movie S5) which elongate under the straining effect of bending deformations in the polarization pattern. A similar chaotic state is also observed for very large contractile activity.

Last, we stress that the results in Sec. \ref{sec:Analysis} are quantitatively unchanged by the frictional interaction between active layer and the surrounding Newtonian fluid, which the diffused-interface model summarized in this Section naturally accounts for. This can be understood by the fact that, at low Reynolds number, such an interaction gives rise to a damping force $\bm{f}_{\rm f}=-\zeta\bm{v}$, with $\zeta$ a drag coefficient, in the active layer \cite{Stone1998}. This, in turn, introduces an additional length scale, i.e. $\ell_{\rm f}=\sqrt{\eta/\zeta}$, reflecting the competition between viscous (i.e. momentum conserving) and frictional dissipation. That is, for distances smaller than $\ell_{\rm f}$, friction does not produce substantial effects and momentum is approximately conserved as in the absence of friction. By contrast, at distances larger than $\ell_{\rm f}$, frictional dissipations takes over and the momentum density decays exponentially. Because the defect-mediated buckling transition discussed here is highly localized around the defect core, friction will not produce any substantial change, unless the coefficient is unrealistically large.

\section{Discussion and Conclusions}

Here, we elucidated how the interplay between topology, geometry and hydrodynamics enables the development of nonplanar features in active liquid crystals, as recently observed experiments on biological and biomimetic systems and, more prominently, in eukaryotic cell layers (Fig. \ref{Fig:fig1} and Refs. \cite{Livshits2017, Braun2018, Maroudas-Sacks2021,Livshits2021}). Whereas the amount of biochemical detail necessary for a thorough account of {\em all} aspects of this phenomenon often results in a complex tangle that hinders fundamental understanding, here, we followed a more generic approach, by focusing on the dynamics at the mesoscopic scale and retaining only the orientational and elastic degrees of freedom, which all realizations of defect-mediated morphogenesis have in common. 

By looking at an axisymmetric surface plagued in the center by a $+1$ disclination, we analytically demonstrated that defective active layers with liquid crystalline order are unstable to out-of-plane deformations and obtained a stability criterion in terms of system size, orientational stiffness, surface tension and active stresses. This mechanism allows defects to serve as topological morphogenes for the formation on nonplanar features, such as domes and protrusion. Such an instability originates from the competition between the focusing of the elastic forces, mediated by defects, and the renormalization of the system’s surface tension by the active flow.
Upon modeling the cell layer as a diffuse active polar interface and turning to computational fluid dynamics, we further investigated the posttransitional regime and constructed an exhaustive phase diagram. At low activity, we recover the results of our analytical approach. Upon increasing the activity, we first find a regime where the layer oscillates periodically between two different buckled states and then a regime where the high activity breaks up the interface leading to continuous droplet nucleation. Last, at very large activity, we see a turbulent regime which is characterized by the chaotic proliferation of protrusions.

Whereas undoubtedly simplified with respect to the seaming endless complexity of living matter, our study contributes to shed light on how developing organisms can take advantage of topology to achieve biological organization from physical mechanisms, with potential application to various biomechanical processes such as morphogenesis, embryogenesis, cancerogenesis and vesicle formation. As mentioned in the introduction, $+1$ disclinations are not the only type of defects encountered in morphogenesis, but nematic  (i.e. $\pm 1/2$; see Fig. \ref{Fig:fig1}A,B) and, in general, $p-$atic defects (e.g. $\pm 1/6$, see Fig. \ref{Fig:fig1}D), are equally common and we will investigate them in the future. Last, while there are several experiments observing the role defects play in the formation of protrusions, we are not aware of any study investigating the other regimes, particularly droplet nucleation. Naturally, the question occurs whether these other regimes also play a role in biological systems.

\section*{Materials and Methods}
\label{sec:mat&met}

\subsection*{Experimental setup}
Parental MDCK GII cells (provided by M. Gloerich, UMC Utrecht) were grown on non-coated coverslips until tissue buckling. F-actin, E-cadherin and nuclei were stained on fixed samples. Samples were imaged at high resolution on a spinning disk confocal microscopy setup. 
Cell boundaries were identified from a maximum intensity projection of a z-stack of the top part of the dome. Fiji software was used for the orthogonal view. 3D reconstructions were done by Imaris Viewer 9.7.0 and $z$ directions were corrected for spherical aberration and axial distortion~\cite{Diel2020}. See the SI for details.

\subsection*{Numerical simulations}
The dynamical fields of the model used for simulations are the incompressible flow field $\bm{v}=\bm{v}(\bm{r},t)$ ($\nabla \cdot \bm{v}=0$) and the polarization field $\bm{P}=\bm{P}(\bm{r},t)$, which is confined at the interface between two isotropic fluids, whose relative concentration is encoded in the scalar phase field $\phi=\phi(\bm{r},t)$. Note that in this section $\bm{P}$ and $\bm{v}$ are three-dimensional vectors, and $\nabla$ now denotes differentiation in $\mathbb{R}^3$.
The equilibrium is defined by a generalized Landau--de Gennes functional~\cite{Carenza22065,carenza2020_physA}:
\begin{multline}
\mathcal{F}\left[\phi,\bm{P}\right] 
= \int {\rm d}V\,\left[ -\dfrac{a}{2}\,\phi^2 +\dfrac{a}{4}\,\phi^4 + \dfrac{k_\phi}{2}\,(\nabla \phi)^2 \right. \\ \left.
+ A_0 \left( \dfrac{\psi}{2}\,|\bm{P}|^2 + \dfrac{1}{4}\,|\bm{P}|^4 \right)
+ \dfrac{\kF}{2}\,|\nabla \bm{P}|^2 + \dfrac{\beta}{2}\,(\bm{P}\cdot \nabla \phi)^2
 \right]\;,
\label{eq:FreeEnergy_sim}
\end{multline}
where the constants $a$ and $k_\phi$ are model parameters related to the surface tension and the width of the interface by $\gamma = \sqrt{8ak_\phi/9}$ and $\xi=\sqrt{2k_\phi/a}$, respectively.
To confine the polar liquid crystal at the $\phi$-interface, the parameter $\psi=\psi(\nabla \phi)$ is chosen to be a function of $\nabla\phi$ such that $\psi=-1$ if $|\nabla \phi|$ is larger than a suitable threshold and $0$ otherwise. In addition, the coupling between $\bm{P}$ and $\nabla \phi$ ensures tangential anchoring of the liquid crystal for $\beta > 0$, so that the polarization field actually lays on the interface. The bulk constant $A_0$ fixes the coherence length of the liquid crystal $ \ell_{\rm c}=\sqrt{\kappa_F/A_0}$, which controls how fast the order parameter $\bm{P}$ drops to zero from its equilibrium value $|\bm{P}|=1$ in proximity of a topological defect.
Note that in the limit of vanishing interfacial width, $\xi \rightarrow 0$, such a phase-field model can be mapped on the analytical model of Sec.~\ref{sec:model} (see e.g. Refs. \cite{Liu2003, Ruske2021}).
The dynamics of the system  is governed by the following set of partial differential equations:
\begin{subequations}\label{eq:hydrodynamics_3d}
\begin{gather}
(\partial_t + \bm{v} \cdot \nabla ) \bm{v} = \nabla \cdot \left(\bm{\sigma}_{\rm p} + \bm{\sigma}_{\rm a} \right)  \;,  \\[7pt]
(\partial_t + \bm{v} \cdot \nabla ) \bm{P} =-\bm{\omega} \cdot \bm{P} + \lambda \bm{u} \cdot \bm{P} + \Gamma \bm{h} \;, \\
\partial_t \phi + \nabla \cdot (\phi \bm{v}) = \nabla \cdot \left( \mu \nabla \dfrac{\delta \mathcal{F}}{\delta \phi} \right) \;.
\end{gather}
\end{subequations}
Equation (\ref{eq:hydrodynamics_3d}a) is the Navier-Stokes equation which rules the hydrodynamics of the system in the full three-dimensional space. Here, the stress tensor has been divided in a passive and an active part. The former includes, in turn, three terms: a hydrodynamic, an elastic and a phase-field contribution $\bm{\sigma}_{\rm p} = \bm{\sigma}_\text{h} + \bm{\sigma}_\text{e}+\bm{\sigma}_\text{i}$, whose explicit expressions are given by 
\begin{gather*}
\bm{\sigma}_{\rm h} = -P_\text{h} \mathbb{1}+\eta\left[\nabla\bm{v}+(\nabla\bm{v})^{\rm T}\right]\;,\\[3pt]
\bm{\sigma}_{\rm e} = \frac{1}{2}\,(\bm{P}\bm{h}-\bm{h}\bm{P})-\frac{\lambda}{2}\,(\bm{P}\bm{h}+\bm{h}\bm{P})-\kappa_{\rm F}\nabla\bm{P}\cdot(\nabla\bm{P})^{\rm T}\;,\\
\bm{\sigma}_{\rm i} = \left(f-\phi\,\frac{\delta\mathcal{F}}{\delta\phi}\right)\mathbb{1}-k_{\phi}\nabla\phi\,\nabla\phi-\beta \bm{P}\nabla\phi\;.
\end{gather*}
Here, $P_\text{h}$ is the hydrodynamic pressure, $\eta$ is the shear viscosity, $\lambda$ is the flow-alignment parameter, and $\Gamma^{-1}$ is the rotational viscosity. $f$ denotes the free energy density, such that $\mathcal{F}=\int {\rm d}V\,f$, $\bm{h} = - \delta \mathcal{F}/\delta \bm{P}$ is the molecular field and $\bm{u}=[(\nabla\bm{v})+(\nabla\bm{v})^{\rm T}]/2$ and $\bm{\omega}=[(\nabla\bm{v})-(\nabla\bm{v})^{\rm T}]/2$ are the strain rate and vorticity tensor, respectively. The three-dimensional active stress tensor $\bm{\sigma}_{\rm a}$ is given by
\begin{equation}
\bm{\sigma}_{\rm a} = \alpha\left(\bm{P}\bm{P}-\frac{1}{3}\,|\bm{P}|^{2}\mathbb{1}\right)\;,	
\end{equation}
Last, Eq. (\ref{eq:hydrodynamics_3d}b) is the Ericksen-Leslie equation for the polarization field $\bm{P}$ in three dimensions, whereas the dynamics of the concentration field is governed by the advection-diffusion equation, Eq. (\ref{eq:hydrodynamics_3d}c), with $\mu$ the mobility parameter.

The dynamical equations have been integrated by means of a  hybrid lattice Boltzmann (LB) method~\cite{carenza2019}, where the hydrodynamics is solved through a \emph{predictor-corrector} LB algorithm, while the dynamics of the order parameter has been treated with a finite-difference approach implementing a first-order upwind scheme and fourth-order accurate stencil for the computation of spacial derivatives.  The hydrodynamics was integrated in a cylindrical geometry with radius $R=32\ell_{\rm c}$ under soft homeotropic boundary conditions for the polarization field, while the concentration field satisfies neutral wetting boundary conditions at the top and bottom walls, respectively, where  $\phi = \pm 1$. The system is initialized with $\phi = 1$ ($\phi = -1$) in the top (bottom) half of the system and the polarization field in an aster configuration at the interface in the center of the system and $0$ otherwise.

The numerical code has been parallelized by means of Message Passage Interface (MPI) by dividing the computational domain in slices and
by implementing the ghost-cell method to compute derivatives on the boundary of the computational subdomains.
Runs have been performed using $64$ CPUs on a computational box size of $96$ by $96$ by $256$, for at least $10^6$ LB iterations (corresponding to $\sim 35$ days of CPU time on Intel Xeon 8160 processors). A mapping between simulation and physical units is provided in the SI.

\section*{Author contributions}{L.A.H., L.N.C., and L.G. designed and performed research, analyzed data and wrote the paper. L.A.H. and L.G. developed the analytical model, L.N.C. the simulations. J.E. performed MDCK GII experiments and analysis.}

\section*{Acknowledgements}
This work is supported by the Netherlands Organization for Scientific Research (NWO/OCW), as part of the Vidi scheme (L.A.H. and L.G.), and by the  European Union via the ERC-CoGgrant HexaTissue (L.N.C. and L.G.). 
Part of this work was carried out on the Dutch National e-Infrastructure with the support of SURF through the grant 2021.028 for computational time (L.A.H., L.N.C., and L.G.).
J.E. and L.G. acknowledge M. Gloerich, UMC Utrecht, for providing us the MDCK cells. L.A.H. thanks Jose A. Santiago for helpful discussions.

\subsection*{Data and materials availability} All data needed to evaluate the conclusions in the paper are present in the paper and/or the Supplementary Materials.

\subsection*{Competing Interests} All authors declare that they have no competing interests.

\bibliography{Biblio.bib}

\begin{thebibliography}{57}%
\makeatletter
\providecommand \@ifxundefined [1]{%
 \@ifx{#1\undefined}
}%
\providecommand \@ifnum [1]{%
 \ifnum #1\expandafter \@firstoftwo
 \else \expandafter \@secondoftwo
 \fi
}%
\providecommand \@ifx [1]{%
 \ifx #1\expandafter \@firstoftwo
 \else \expandafter \@secondoftwo
 \fi
}%
\providecommand \natexlab [1]{#1}%
\providecommand \enquote  [1]{``#1''}%
\providecommand \bibnamefont  [1]{#1}%
\providecommand \bibfnamefont [1]{#1}%
\providecommand \citenamefont [1]{#1}%
\providecommand \href@noop [0]{\@secondoftwo}%
\providecommand \href [0]{\begingroup \@sanitize@url \@href}%
\providecommand \@href[1]{\@@startlink{#1}\@@href}%
\providecommand \@@href[1]{\endgroup#1\@@endlink}%
\providecommand \@sanitize@url [0]{\catcode `\\12\catcode `\$12\catcode
  `\&12\catcode `\#12\catcode `\^12\catcode `\_12\catcode `\%12\relax}%
\providecommand \@@startlink[1]{}%
\providecommand \@@endlink[0]{}%
\providecommand \url  [0]{\begingroup\@sanitize@url \@url }%
\providecommand \@url [1]{\endgroup\@href {#1}{\urlprefix }}%
\providecommand \urlprefix  [0]{URL }%
\providecommand \Eprint [0]{\href }%
\providecommand \doibase [0]{https://doi.org/}%
\providecommand \selectlanguage [0]{\@gobble}%
\providecommand \bibinfo  [0]{\@secondoftwo}%
\providecommand \bibfield  [0]{\@secondoftwo}%
\providecommand \translation [1]{[#1]}%
\providecommand \BibitemOpen [0]{}%
\providecommand \bibitemStop [0]{}%
\providecommand \bibitemNoStop [0]{.\EOS\space}%
\providecommand \EOS [0]{\spacefactor3000\relax}%
\providecommand \BibitemShut  [1]{\csname bibitem#1\endcsname}%
\let\auto@bib@innerbib\@empty
\bibitem [{\citenamefont {Conte}\ \emph {et~al.}(2012)\citenamefont {Conte},
  \citenamefont {Ulrich}, \citenamefont {Baum}, \citenamefont {Muñoz},
  \citenamefont {Veldhuis}, \citenamefont {Brodland},\ and\ \citenamefont
  {Miodownik}}]{Conte2012}%
  \BibitemOpen
  \bibfield  {author} {\bibinfo {author} {\bibfnamefont {V.}~\bibnamefont
  {Conte}}, \bibinfo {author} {\bibfnamefont {F.}~\bibnamefont {Ulrich}},
  \bibinfo {author} {\bibfnamefont {B.}~\bibnamefont {Baum}}, \bibinfo {author}
  {\bibfnamefont {J.}~\bibnamefont {Muñoz}}, \bibinfo {author} {\bibfnamefont
  {J.}~\bibnamefont {Veldhuis}}, \bibinfo {author} {\bibfnamefont
  {W.}~\bibnamefont {Brodland}},\ and\ \bibinfo {author} {\bibfnamefont
  {M.}~\bibnamefont {Miodownik}},\ }\bibfield  {title} {\bibinfo {title} {A
  {Biomechanical} {Analysis} of {Ventral} {Furrow} {Formation} in the
  {Drosophila} {Melanogaster} {Embryo}},\ }\href
  {https://doi.org/10.1371/journal.pone.0034473} {\bibfield  {journal}
  {\bibinfo  {journal} {PLOS ONE}\ }\textbf {\bibinfo {volume} {7}},\ \bibinfo
  {pages} {e34473} (\bibinfo {year} {2012})}\BibitemShut {NoStop}%
\bibitem [{\citenamefont {Nelson}(2016)}]{Nelson2016}%
  \BibitemOpen
  \bibfield  {author} {\bibinfo {author} {\bibfnamefont {C.~M.}\ \bibnamefont
  {Nelson}},\ }\bibfield  {title} {\bibinfo {title} {On buckling
  morphogenesis},\ }\href {https://doi.org/10.1115/1.4032128} {\bibfield
  {journal} {\bibinfo  {journal} {J. Biomech. Eng.}\ }\textbf {\bibinfo
  {volume} {138}},\ \bibinfo {pages} {021005} (\bibinfo {year}
  {2016})}\BibitemShut {NoStop}%
\bibitem [{\citenamefont {Marchetti}\ \emph {et~al.}(2013)\citenamefont
  {Marchetti}, \citenamefont {Joanny}, \citenamefont {Ramaswamy}, \citenamefont
  {Liverpool}, \citenamefont {Prost}, \citenamefont {Rao},\ and\ \citenamefont
  {Simha}}]{Marchetti2013}%
  \BibitemOpen
  \bibfield  {author} {\bibinfo {author} {\bibfnamefont {M.~C.}\ \bibnamefont
  {Marchetti}}, \bibinfo {author} {\bibfnamefont {J.~F.}\ \bibnamefont
  {Joanny}}, \bibinfo {author} {\bibfnamefont {S.}~\bibnamefont {Ramaswamy}},
  \bibinfo {author} {\bibfnamefont {T.~B.}\ \bibnamefont {Liverpool}}, \bibinfo
  {author} {\bibfnamefont {J.}~\bibnamefont {Prost}}, \bibinfo {author}
  {\bibfnamefont {M.}~\bibnamefont {Rao}},\ and\ \bibinfo {author}
  {\bibfnamefont {R.~A.}\ \bibnamefont {Simha}},\ }\bibfield  {title} {\bibinfo
  {title} {Hydrodynamics of soft active matter},\ }\href
  {https://doi.org/10.1103/RevModPhys.85.1143} {\bibfield  {journal} {\bibinfo
  {journal} {Rev. Mod. Phys.}\ }\textbf {\bibinfo {volume} {85}},\ \bibinfo
  {pages} {1143} (\bibinfo {year} {2013})}\BibitemShut {NoStop}%
\bibitem [{\citenamefont {Metselaar}\ \emph {et~al.}(2019)\citenamefont
  {Metselaar}, \citenamefont {Yeomans},\ and\ \citenamefont
  {Doostmohammadi}}]{Metselaar2019}%
  \BibitemOpen
  \bibfield  {author} {\bibinfo {author} {\bibfnamefont {L.}~\bibnamefont
  {Metselaar}}, \bibinfo {author} {\bibfnamefont {J.~M.}\ \bibnamefont
  {Yeomans}},\ and\ \bibinfo {author} {\bibfnamefont {A.}~\bibnamefont
  {Doostmohammadi}},\ }\bibfield  {title} {\bibinfo {title} {Topology and
  {Morphology} of {Self}-{Deforming} {Active} {Shells}},\ }\href
  {https://doi.org/10.1103/PhysRevLett.123.208001} {\bibfield  {journal}
  {\bibinfo  {journal} {Phys. Rev. Lett.}\ }\textbf {\bibinfo {volume} {123}},\
  \bibinfo {pages} {208001} (\bibinfo {year} {2019})}\BibitemShut {NoStop}%
\bibitem [{\citenamefont {Mietke}\ \emph
  {et~al.}(2019{\natexlab{a}})\citenamefont {Mietke}, \citenamefont
  {Jülicher},\ and\ \citenamefont {Sbalzarini}}]{Mietke2019a}%
  \BibitemOpen
  \bibfield  {author} {\bibinfo {author} {\bibfnamefont {A.}~\bibnamefont
  {Mietke}}, \bibinfo {author} {\bibfnamefont {F.}~\bibnamefont {Jülicher}},\
  and\ \bibinfo {author} {\bibfnamefont {I.~F.}\ \bibnamefont {Sbalzarini}},\
  }\bibfield  {title} {\bibinfo {title} {Self-organized shape dynamics of
  active surfaces},\ }\href {https://doi.org/10.1073/pnas.1810896115}
  {\bibfield  {journal} {\bibinfo  {journal} {Proc. Natl. Acad. Sci. U.S.A.}\
  }\textbf {\bibinfo {volume} {116}},\ \bibinfo {pages} {29} (\bibinfo {year}
  {2019}{\natexlab{a}})}\BibitemShut {NoStop}%
\bibitem [{\citenamefont {Wyatt}\ \emph {et~al.}(2020)\citenamefont {Wyatt},
  \citenamefont {Fouchard}, \citenamefont {Lisica}, \citenamefont
  {Khalilgharibi}, \citenamefont {Baum}, \citenamefont {Recho}, \citenamefont
  {Kabla},\ and\ \citenamefont {Charras}}]{Wyatt2020}%
  \BibitemOpen
  \bibfield  {author} {\bibinfo {author} {\bibfnamefont {T.~P.~J.}\
  \bibnamefont {Wyatt}}, \bibinfo {author} {\bibfnamefont {J.}~\bibnamefont
  {Fouchard}}, \bibinfo {author} {\bibfnamefont {A.}~\bibnamefont {Lisica}},
  \bibinfo {author} {\bibfnamefont {N.}~\bibnamefont {Khalilgharibi}}, \bibinfo
  {author} {\bibfnamefont {B.}~\bibnamefont {Baum}}, \bibinfo {author}
  {\bibfnamefont {P.}~\bibnamefont {Recho}}, \bibinfo {author} {\bibfnamefont
  {A.~J.}\ \bibnamefont {Kabla}},\ and\ \bibinfo {author} {\bibfnamefont
  {G.~T.}\ \bibnamefont {Charras}},\ }\bibfield  {title} {\bibinfo {title}
  {Actomyosin controls planarity and folding of epithelia in response to
  compression},\ }\href {https://doi.org/10.1038/s41563-019-0461-x} {\bibfield
  {journal} {\bibinfo  {journal} {Nat. Mater.}\ }\textbf {\bibinfo {volume}
  {19}},\ \bibinfo {pages} {109} (\bibinfo {year} {2020})}\BibitemShut
  {NoStop}%
\bibitem [{\citenamefont {Al-Izzi}\ and\ \citenamefont
  {Morris}(2021)}]{Al-Izzi2021}%
  \BibitemOpen
  \bibfield  {author} {\bibinfo {author} {\bibfnamefont {S.~C.}\ \bibnamefont
  {Al-Izzi}}\ and\ \bibinfo {author} {\bibfnamefont {R.~G.}\ \bibnamefont
  {Morris}},\ }\bibfield  {title} {\bibinfo {title} {Active {Flows} and
  {Deformable} {Surfaces} in {Development}},\ }\href
  {http://arxiv.org/abs/2103.12264} {\bibfield  {journal} {\bibinfo  {journal}
  {arXiv:2103.12264}\ } (\bibinfo {year} {2021})}\BibitemShut {NoStop}%
\bibitem [{\citenamefont {Keber}\ \emph {et~al.}(2014)\citenamefont {Keber},
  \citenamefont {Loiseau}, \citenamefont {Sanchez}, \citenamefont {DeCamp},
  \citenamefont {Giomi}, \citenamefont {Bowick}, \citenamefont {Marchetti},
  \citenamefont {Dogic},\ and\ \citenamefont {Bausch}}]{Keber2014}%
  \BibitemOpen
  \bibfield  {author} {\bibinfo {author} {\bibfnamefont {F.~C.}\ \bibnamefont
  {Keber}}, \bibinfo {author} {\bibfnamefont {E.}~\bibnamefont {Loiseau}},
  \bibinfo {author} {\bibfnamefont {T.}~\bibnamefont {Sanchez}}, \bibinfo
  {author} {\bibfnamefont {S.~J.}\ \bibnamefont {DeCamp}}, \bibinfo {author}
  {\bibfnamefont {L.}~\bibnamefont {Giomi}}, \bibinfo {author} {\bibfnamefont
  {M.~J.}\ \bibnamefont {Bowick}}, \bibinfo {author} {\bibfnamefont {M.~C.}\
  \bibnamefont {Marchetti}}, \bibinfo {author} {\bibfnamefont {Z.}~\bibnamefont
  {Dogic}},\ and\ \bibinfo {author} {\bibfnamefont {A.~R.}\ \bibnamefont
  {Bausch}},\ }\bibfield  {title} {\bibinfo {title} {Topology and dynamics of
  active nematic vesicles},\ }\href {https://doi.org/10.1126/science.1254784}
  {\bibfield  {journal} {\bibinfo  {journal} {Science}\ }\textbf {\bibinfo
  {volume} {345}},\ \bibinfo {pages} {1135} (\bibinfo {year}
  {2014})}\BibitemShut {NoStop}%
\bibitem [{\citenamefont {Livshits}\ \emph {et~al.}(2017)\citenamefont
  {Livshits}, \citenamefont {Shani-Zerbib}, \citenamefont {Maroudas-Sacks},
  \citenamefont {Braun},\ and\ \citenamefont {Keren}}]{Livshits2017}%
  \BibitemOpen
  \bibfield  {author} {\bibinfo {author} {\bibfnamefont {A.}~\bibnamefont
  {Livshits}}, \bibinfo {author} {\bibfnamefont {L.}~\bibnamefont
  {Shani-Zerbib}}, \bibinfo {author} {\bibfnamefont {Y.}~\bibnamefont
  {Maroudas-Sacks}}, \bibinfo {author} {\bibfnamefont {E.}~\bibnamefont
  {Braun}},\ and\ \bibinfo {author} {\bibfnamefont {K.}~\bibnamefont {Keren}},\
  }\bibfield  {title} {\bibinfo {title} {Structural {Inheritance} of the
  {Actin} {Cytoskeletal} {Organization} {Determines} the {Body} {Axis} in
  {Regenerating} {Hydra}},\ }\href
  {https://doi.org/10.1016/j.celrep.2017.01.036} {\bibfield  {journal}
  {\bibinfo  {journal} {Cell Rep.}\ }\textbf {\bibinfo {volume} {18}},\
  \bibinfo {pages} {1410} (\bibinfo {year} {2017})}\BibitemShut {NoStop}%
\bibitem [{\citenamefont {Braun}\ and\ \citenamefont
  {Keren}(2018)}]{Braun2018}%
  \BibitemOpen
  \bibfield  {author} {\bibinfo {author} {\bibfnamefont {E.}~\bibnamefont
  {Braun}}\ and\ \bibinfo {author} {\bibfnamefont {K.}~\bibnamefont {Keren}},\
  }\bibfield  {title} {\bibinfo {title} {\textit{{Hydra}} regeneration:
  {Closing} the loop with mechanical processes in morphogenesis},\ }\href
  {https://doi.org/10.1002/bies.201700204} {\bibfield  {journal} {\bibinfo
  {journal} {Bioessays}\ }\textbf {\bibinfo {volume} {40}},\ \bibinfo {pages}
  {1700204} (\bibinfo {year} {2018})}\BibitemShut {NoStop}%
\bibitem [{\citenamefont {Maroudas-Sacks}\ \emph {et~al.}(2021)\citenamefont
  {Maroudas-Sacks}, \citenamefont {Garion}, \citenamefont {Shani-Zerbib},
  \citenamefont {Livshits}, \citenamefont {Braun},\ and\ \citenamefont
  {Keren}}]{Maroudas-Sacks2021}%
  \BibitemOpen
  \bibfield  {author} {\bibinfo {author} {\bibfnamefont {Y.}~\bibnamefont
  {Maroudas-Sacks}}, \bibinfo {author} {\bibfnamefont {L.}~\bibnamefont
  {Garion}}, \bibinfo {author} {\bibfnamefont {L.}~\bibnamefont
  {Shani-Zerbib}}, \bibinfo {author} {\bibfnamefont {A.}~\bibnamefont
  {Livshits}}, \bibinfo {author} {\bibfnamefont {E.}~\bibnamefont {Braun}},\
  and\ \bibinfo {author} {\bibfnamefont {K.}~\bibnamefont {Keren}},\ }\bibfield
   {title} {\bibinfo {title} {Topological defects in the nematic order of actin
  fibres as organization centers of \textit{{Hydra}} morphogenesis},\ }\href
  {https://doi.org/10.1038/s41567-020-01083-1} {\bibfield  {journal} {\bibinfo
  {journal} {Nat. Phys.}\ }\textbf {\bibinfo {volume} {17}},\ \bibinfo {pages}
  {251} (\bibinfo {year} {2021})}\BibitemShut {NoStop}%
\bibitem [{\citenamefont {Livshits}\ \emph {et~al.}(2021)\citenamefont
  {Livshits}, \citenamefont {Garion}, \citenamefont {Maroudas-Sacks},
  \citenamefont {Shani-Zerbib}, \citenamefont {Keren},\ and\ \citenamefont
  {Braun}}]{Livshits2021}%
  \BibitemOpen
  \bibfield  {author} {\bibinfo {author} {\bibfnamefont {A.}~\bibnamefont
  {Livshits}}, \bibinfo {author} {\bibfnamefont {L.}~\bibnamefont {Garion}},
  \bibinfo {author} {\bibfnamefont {Y.}~\bibnamefont {Maroudas-Sacks}},
  \bibinfo {author} {\bibfnamefont {L.}~\bibnamefont {Shani-Zerbib}}, \bibinfo
  {author} {\bibfnamefont {K.}~\bibnamefont {Keren}},\ and\ \bibinfo {author}
  {\bibfnamefont {E.}~\bibnamefont {Braun}},\ }\bibfield  {title} {\bibinfo
  {title} {Plasticity of body axis polarity in \textit{{Hydra}} regeneration
  under constraints},\ }\href
  {http://biorxiv.org/lookup/doi/10.1101/2021.02.04.429818} {\bibfield
  {journal} {\bibinfo  {journal} {bioRxiv 2021.02.04.429818}\ } (\bibinfo
  {year} {2021})}\BibitemShut {NoStop}%
\bibitem [{\citenamefont {Ruske}\ and\ \citenamefont
  {Yeomans}(2021)}]{Ruske2021}%
  \BibitemOpen
  \bibfield  {author} {\bibinfo {author} {\bibfnamefont {L.~J.}\ \bibnamefont
  {Ruske}}\ and\ \bibinfo {author} {\bibfnamefont {J.~M.}\ \bibnamefont
  {Yeomans}},\ }\bibfield  {title} {\bibinfo {title} {Morphology of active
  deformable {3D} droplets},\ }\href
  {https://doi.org/10.1103/PhysRevX.11.021001} {\bibfield  {journal} {\bibinfo
  {journal} {Phys. Rev. X}\ }\textbf {\bibinfo {volume} {11}},\ \bibinfo
  {pages} {021001} (\bibinfo {year} {2021})}\BibitemShut {NoStop}%
\bibitem [{\citenamefont {Vafa}\ and\ \citenamefont
  {Mahadevan}(2021)}]{Vafa2021}%
  \BibitemOpen
  \bibfield  {author} {\bibinfo {author} {\bibfnamefont {F.}~\bibnamefont
  {Vafa}}\ and\ \bibinfo {author} {\bibfnamefont {L.}~\bibnamefont
  {Mahadevan}},\ }\bibfield  {title} {\bibinfo {title} {Active nematic defects
  and epithelial morphogenesis},\ }\href {https://arxiv.org/abs/2105.01067}
  {\bibfield  {journal} {\bibinfo  {journal} {arXiv:2105.01067}\ } (\bibinfo
  {year} {2021})}\BibitemShut {NoStop}%
\bibitem [{\citenamefont {Saw}\ \emph {et~al.}(2017)\citenamefont {Saw},
  \citenamefont {Doostmohammadi}, \citenamefont {Nier}, \citenamefont
  {Kocgozlu}, \citenamefont {Thampi}, \citenamefont {Toyama}, \citenamefont
  {Marcq}, \citenamefont {Lim}, \citenamefont {Yeomans},\ and\ \citenamefont
  {Ladoux}}]{Saw2017}%
  \BibitemOpen
  \bibfield  {author} {\bibinfo {author} {\bibfnamefont {T.~B.}\ \bibnamefont
  {Saw}}, \bibinfo {author} {\bibfnamefont {A.}~\bibnamefont {Doostmohammadi}},
  \bibinfo {author} {\bibfnamefont {V.}~\bibnamefont {Nier}}, \bibinfo {author}
  {\bibfnamefont {L.}~\bibnamefont {Kocgozlu}}, \bibinfo {author}
  {\bibfnamefont {S.}~\bibnamefont {Thampi}}, \bibinfo {author} {\bibfnamefont
  {Y.}~\bibnamefont {Toyama}}, \bibinfo {author} {\bibfnamefont
  {P.}~\bibnamefont {Marcq}}, \bibinfo {author} {\bibfnamefont {C.~T.}\
  \bibnamefont {Lim}}, \bibinfo {author} {\bibfnamefont {J.~M.}\ \bibnamefont
  {Yeomans}},\ and\ \bibinfo {author} {\bibfnamefont {B.}~\bibnamefont
  {Ladoux}},\ }\bibfield  {title} {\bibinfo {title} {Topological defects in
  epithelia govern cell death and extrusion},\ }\href
  {https://doi.org/10.1038/nature21718} {\bibfield  {journal} {\bibinfo
  {journal} {Nature}\ }\textbf {\bibinfo {volume} {544}},\ \bibinfo {pages}
  {212} (\bibinfo {year} {2017})}\BibitemShut {NoStop}%
\bibitem [{\citenamefont {Loewe}\ \emph {et~al.}(2020)\citenamefont {Loewe},
  \citenamefont {Chiang}, \citenamefont {Marenduzzo},\ and\ \citenamefont
  {Marchetti}}]{Loewe2020}%
  \BibitemOpen
  \bibfield  {author} {\bibinfo {author} {\bibfnamefont {B.}~\bibnamefont
  {Loewe}}, \bibinfo {author} {\bibfnamefont {M.}~\bibnamefont {Chiang}},
  \bibinfo {author} {\bibfnamefont {D.}~\bibnamefont {Marenduzzo}},\ and\
  \bibinfo {author} {\bibfnamefont {M.~C.}\ \bibnamefont {Marchetti}},\
  }\bibfield  {title} {\bibinfo {title} {Solid-liquid transition of deformable
  and overlapping active particles},\ }\href
  {https://doi.org/10.1103/PhysRevLett.125.038003} {\bibfield  {journal}
  {\bibinfo  {journal} {Phys. Rev. Lett.}\ }\textbf {\bibinfo {volume} {125}},\
  \bibinfo {pages} {038003} (\bibinfo {year} {2020})}\BibitemShut {NoStop}%
\bibitem [{\citenamefont {Monfared}\ \emph {et~al.}(2021)\citenamefont
  {Monfared}, \citenamefont {Ravichandran}, \citenamefont {Andrade},\ and\
  \citenamefont {Doostmohammadi}}]{Monfared2021}%
  \BibitemOpen
  \bibfield  {author} {\bibinfo {author} {\bibfnamefont {S.}~\bibnamefont
  {Monfared}}, \bibinfo {author} {\bibfnamefont {G.}~\bibnamefont
  {Ravichandran}}, \bibinfo {author} {\bibfnamefont {J.~E.}\ \bibnamefont
  {Andrade}},\ and\ \bibinfo {author} {\bibfnamefont {A.}~\bibnamefont
  {Doostmohammadi}},\ }\bibfield  {title} {\bibinfo {title} {Mechanics of live
  cell elimination},\ }\href {https://arxiv.org/abs/2108.07657} {\bibfield
  {journal} {\bibinfo  {journal} {arXiv:2108.07657}\ } (\bibinfo {year}
  {2021})}\BibitemShut {NoStop}%
\bibitem [{\citenamefont {Guillamat}\ \emph {et~al.}(2020)\citenamefont
  {Guillamat}, \citenamefont {Blanch-Mercader}, \citenamefont {Kruse},\ and\
  \citenamefont {Roux}}]{Guillamat2020}%
  \BibitemOpen
  \bibfield  {author} {\bibinfo {author} {\bibfnamefont {P.}~\bibnamefont
  {Guillamat}}, \bibinfo {author} {\bibfnamefont {C.}~\bibnamefont
  {Blanch-Mercader}}, \bibinfo {author} {\bibfnamefont {K.}~\bibnamefont
  {Kruse}},\ and\ \bibinfo {author} {\bibfnamefont {A.}~\bibnamefont {Roux}},\
  }\bibfield  {title} {\bibinfo {title} {Integer topological defects organize
  stresses driving tissue morphogenesis},\ }\href
  {https://www.biorxiv.org/content/10.1101/2020.06.02.129262v1} {\bibfield
  {journal} {\bibinfo  {journal} {bioRxiv 2020.06.02.129262}\ } (\bibinfo
  {year} {2020})}\BibitemShut {NoStop}%
\bibitem [{\citenamefont {Popowicz}\ \emph {et~al.}(1986)\citenamefont
  {Popowicz}, \citenamefont {Kurzyca},\ and\ \citenamefont
  {Popowicz}}]{Popowicz1986}%
  \BibitemOpen
  \bibfield  {author} {\bibinfo {author} {\bibfnamefont {P.}~\bibnamefont
  {Popowicz}}, \bibinfo {author} {\bibfnamefont {J.}~\bibnamefont {Kurzyca}},\
  and\ \bibinfo {author} {\bibfnamefont {S.}~\bibnamefont {Popowicz}},\
  }\bibfield  {title} {\bibinfo {title} {``{D}ome-curve'' — three size
  classes of domes of {MDCK} epithelial monolayer},\ }\href
  {https://doi.org/https://doi.org/10.1016/S0232-1513(86)80010-X} {\bibfield
  {journal} {\bibinfo  {journal} {Exp. Pathol.}\ }\textbf {\bibinfo {volume}
  {29}},\ \bibinfo {pages} {147} (\bibinfo {year} {1986})}\BibitemShut
  {NoStop}%
\bibitem [{\citenamefont {Arslan}\ \emph {et~al.}(2021)\citenamefont {Arslan},
  \citenamefont {Eckert}, \citenamefont {Schmidt},\ and\ \citenamefont
  {Heisenberg}}]{Arslan2021}%
  \BibitemOpen
  \bibfield  {author} {\bibinfo {author} {\bibfnamefont {F.~N.}\ \bibnamefont
  {Arslan}}, \bibinfo {author} {\bibfnamefont {J.}~\bibnamefont {Eckert}},
  \bibinfo {author} {\bibfnamefont {T.}~\bibnamefont {Schmidt}},\ and\ \bibinfo
  {author} {\bibfnamefont {C.-P.}\ \bibnamefont {Heisenberg}},\ }\bibfield
  {title} {\bibinfo {title} {Holding it together: {When} cadherin meets
  cadherin},\ }\bibfield  {journal} {\bibinfo  {journal} {Biophys. J.}\ }\href
  {https://doi.org/10.1016/j.bpj.2021.03.025} {10.1016/j.bpj.2021.03.025}
  (\bibinfo {year} {2021})\BibitemShut {NoStop}%
\bibitem [{\citenamefont {Cereijido}\ \emph {et~al.}(1978)\citenamefont
  {Cereijido}, \citenamefont {Robbins}, \citenamefont {Dolan}, \citenamefont
  {Rotunno},\ and\ \citenamefont {Sabatini}}]{Cereijido1978}%
  \BibitemOpen
  \bibfield  {author} {\bibinfo {author} {\bibfnamefont {M.}~\bibnamefont
  {Cereijido}}, \bibinfo {author} {\bibfnamefont {E.}~\bibnamefont {Robbins}},
  \bibinfo {author} {\bibfnamefont {W.}~\bibnamefont {Dolan}}, \bibinfo
  {author} {\bibfnamefont {C.}~\bibnamefont {Rotunno}},\ and\ \bibinfo {author}
  {\bibfnamefont {D.}~\bibnamefont {Sabatini}},\ }\bibfield  {title} {\bibinfo
  {title} {{Polarized monolayers formed by epithelial cells on a permeable and
  translucent support}},\ }\href {https://doi.org/10.1083/jcb.77.3.853}
  {\bibfield  {journal} {\bibinfo  {journal} {J. Cell Biol.}\ }\textbf
  {\bibinfo {volume} {77}},\ \bibinfo {pages} {853} (\bibinfo {year}
  {1978})}\BibitemShut {NoStop}%
\bibitem [{\citenamefont {Latorre}\ \emph {et~al.}(2018)\citenamefont
  {Latorre}, \citenamefont {Kale}, \citenamefont {Casares}, \citenamefont
  {Gómez-González}, \citenamefont {Uroz}, \citenamefont {Valon},
  \citenamefont {Nair}, \citenamefont {Garreta}, \citenamefont {Montserrat},
  \citenamefont {del Campo}, \citenamefont {Ladoux}, \citenamefont {Arroyo},\
  and\ \citenamefont {Trepat}}]{Latorre2018}%
  \BibitemOpen
  \bibfield  {author} {\bibinfo {author} {\bibfnamefont {E.}~\bibnamefont
  {Latorre}}, \bibinfo {author} {\bibfnamefont {S.}~\bibnamefont {Kale}},
  \bibinfo {author} {\bibfnamefont {L.}~\bibnamefont {Casares}}, \bibinfo
  {author} {\bibfnamefont {M.}~\bibnamefont {Gómez-González}}, \bibinfo
  {author} {\bibfnamefont {M.}~\bibnamefont {Uroz}}, \bibinfo {author}
  {\bibfnamefont {L.}~\bibnamefont {Valon}}, \bibinfo {author} {\bibfnamefont
  {R.~V.}\ \bibnamefont {Nair}}, \bibinfo {author} {\bibfnamefont
  {E.}~\bibnamefont {Garreta}}, \bibinfo {author} {\bibfnamefont
  {N.}~\bibnamefont {Montserrat}}, \bibinfo {author} {\bibfnamefont
  {A.}~\bibnamefont {del Campo}}, \bibinfo {author} {\bibfnamefont
  {B.}~\bibnamefont {Ladoux}}, \bibinfo {author} {\bibfnamefont
  {M.}~\bibnamefont {Arroyo}},\ and\ \bibinfo {author} {\bibfnamefont
  {X.}~\bibnamefont {Trepat}},\ }\bibfield  {title} {\bibinfo {title} {Active
  superelasticity in three-dimensional epithelia of controlled shape},\ }\href
  {https://doi.org/10.1038/s41586-018-0671-4} {\bibfield  {journal} {\bibinfo
  {journal} {Nature}\ }\textbf {\bibinfo {volume} {563}},\ \bibinfo {pages}
  {203} (\bibinfo {year} {2018})}\BibitemShut {NoStop}%
\bibitem [{\citenamefont {Deserno}(2015)}]{Deserno2015}%
  \BibitemOpen
  \bibfield  {author} {\bibinfo {author} {\bibfnamefont {M.}~\bibnamefont
  {Deserno}},\ }\bibfield  {title} {\bibinfo {title} {Fluid lipid membranes:
  {From} differential geometry to curvature stresses},\ }\href
  {https://doi.org/10.1016/j.chemphyslip.2014.05.001} {\bibfield  {journal}
  {\bibinfo  {journal} {Chem. Phys. Lipids}\ }\textbf {\bibinfo {volume}
  {185}},\ \bibinfo {pages} {11} (\bibinfo {year} {2015})}\BibitemShut
  {NoStop}%
\bibitem [{\citenamefont {Helfrich}(1973)}]{Helfrich1973}%
  \BibitemOpen
  \bibfield  {author} {\bibinfo {author} {\bibfnamefont {W.}~\bibnamefont
  {Helfrich}},\ }\bibfield  {title} {\bibinfo {title} {Elastic properties of
  lipid bilayers: {Theory} and possible experiments},\ }\href@noop {}
  {\bibfield  {journal} {\bibinfo  {journal} {Z. Naturforschung C}\ }\textbf
  {\bibinfo {volume} {28}},\ \bibinfo {pages} {693} (\bibinfo {year}
  {1973})}\BibitemShut {NoStop}%
\bibitem [{\citenamefont {Chaikin}\ and\ \citenamefont
  {Lubensky}(1995)}]{Chaikin1995}%
  \BibitemOpen
  \bibfield  {author} {\bibinfo {author} {\bibfnamefont {P.~M.}\ \bibnamefont
  {Chaikin}}\ and\ \bibinfo {author} {\bibfnamefont {T.~C.}\ \bibnamefont
  {Lubensky}},\ }\href@noop {} {\emph {\bibinfo {title} {Principles of
  Condensed Matter Physics}}}\ (\bibinfo  {publisher} {Cambridge University
  Press},\ \bibinfo {year} {1995})\BibitemShut {NoStop}%
\bibitem [{\citenamefont {Giomi}(2015)}]{Giomi2015}%
  \BibitemOpen
  \bibfield  {author} {\bibinfo {author} {\bibfnamefont {L.}~\bibnamefont
  {Giomi}},\ }\bibfield  {title} {\bibinfo {title} {Geometry and {Topology} of
  {Turbulence} in {Active} {Nematics}},\ }\href
  {https://doi.org/10.1103/PhysRevX.5.031003} {\bibfield  {journal} {\bibinfo
  {journal} {Phys. Rev. X}\ }\textbf {\bibinfo {volume} {5}},\ \bibinfo {pages}
  {031003} (\bibinfo {year} {2015})}\BibitemShut {NoStop}%
\bibitem [{\citenamefont {Pearce}\ \emph {et~al.}(2019)\citenamefont {Pearce},
  \citenamefont {Ellis}, \citenamefont {Fernandez-Nieves},\ and\ \citenamefont
  {Giomi}}]{Pearce2019}%
  \BibitemOpen
  \bibfield  {author} {\bibinfo {author} {\bibfnamefont {D.~J.~G.}\
  \bibnamefont {Pearce}}, \bibinfo {author} {\bibfnamefont {P.~W.}\
  \bibnamefont {Ellis}}, \bibinfo {author} {\bibfnamefont {A.}~\bibnamefont
  {Fernandez-Nieves}},\ and\ \bibinfo {author} {\bibfnamefont {L.}~\bibnamefont
  {Giomi}},\ }\bibfield  {title} {\bibinfo {title} {Geometrical {Control} of
  {Active} {Turbulence} in {Curved} {Topographies}},\ }\href
  {https://doi.org/10.1103/PhysRevLett.122.168002} {\bibfield  {journal}
  {\bibinfo  {journal} {Phys. Rev. Lett.}\ }\textbf {\bibinfo {volume} {122}},\
  \bibinfo {pages} {168002} (\bibinfo {year} {2019})}\BibitemShut {NoStop}%
\bibitem [{\citenamefont {Mietke}\ \emph
  {et~al.}(2019{\natexlab{b}})\citenamefont {Mietke}, \citenamefont {Jemseena},
  \citenamefont {Kumar}, \citenamefont {Sbalzarini},\ and\ \citenamefont
  {Jülicher}}]{Mietke2019b}%
  \BibitemOpen
  \bibfield  {author} {\bibinfo {author} {\bibfnamefont {A.}~\bibnamefont
  {Mietke}}, \bibinfo {author} {\bibfnamefont {V.}~\bibnamefont {Jemseena}},
  \bibinfo {author} {\bibfnamefont {K.~V.}\ \bibnamefont {Kumar}}, \bibinfo
  {author} {\bibfnamefont {I.~F.}\ \bibnamefont {Sbalzarini}},\ and\ \bibinfo
  {author} {\bibfnamefont {F.}~\bibnamefont {Jülicher}},\ }\bibfield  {title}
  {\bibinfo {title} {Minimal {Model} of {Cellular} {Symmetry} {Breaking}},\
  }\href {https://doi.org/10.1103/PhysRevLett.123.188101} {\bibfield  {journal}
  {\bibinfo  {journal} {Phys. Rev. Lett.}\ }\textbf {\bibinfo {volume} {123}},\
  \bibinfo {pages} {188101} (\bibinfo {year} {2019}{\natexlab{b}})}\BibitemShut
  {NoStop}%
\bibitem [{\citenamefont {Santiago}(2018)}]{Santiago2018}%
  \BibitemOpen
  \bibfield  {author} {\bibinfo {author} {\bibfnamefont {J.~A.}\ \bibnamefont
  {Santiago}},\ }\bibfield  {title} {\bibinfo {title} {Stresses in curved
  nematic membranes},\ }\href {https://doi.org/10.1103/PhysRevE.97.052706}
  {\bibfield  {journal} {\bibinfo  {journal} {Phys. Rev. E}\ }\textbf {\bibinfo
  {volume} {97}},\ \bibinfo {pages} {052706} (\bibinfo {year}
  {2018})}\BibitemShut {NoStop}%
\bibitem [{\citenamefont {T\'oth}\ \emph {et~al.}(2002)\citenamefont {T\'oth},
  \citenamefont {Denniston},\ and\ \citenamefont {Yeomans}}]{Toth2002}%
  \BibitemOpen
  \bibfield  {author} {\bibinfo {author} {\bibfnamefont {G.}~\bibnamefont
  {T\'oth}}, \bibinfo {author} {\bibfnamefont {C.}~\bibnamefont {Denniston}},\
  and\ \bibinfo {author} {\bibfnamefont {J.~M.}\ \bibnamefont {Yeomans}},\
  }\bibfield  {title} {\bibinfo {title} {Hydrodynamics of topological defects
  in nematic liquid crystals},\ }\href
  {https://doi.org/10.1103/PhysRevLett.88.105504} {\bibfield  {journal}
  {\bibinfo  {journal} {Phys. Rev. Lett.}\ }\textbf {\bibinfo {volume} {88}},\
  \bibinfo {pages} {105504} (\bibinfo {year} {2002})}\BibitemShut {NoStop}%
\bibitem [{\citenamefont {Bowick}\ \emph {et~al.}(2000)\citenamefont {Bowick},
  \citenamefont {Nelson},\ and\ \citenamefont {Travesset}}]{Bowick2000}%
  \BibitemOpen
  \bibfield  {author} {\bibinfo {author} {\bibfnamefont {M.~J.}\ \bibnamefont
  {Bowick}}, \bibinfo {author} {\bibfnamefont {D.~R.}\ \bibnamefont {Nelson}},\
  and\ \bibinfo {author} {\bibfnamefont {A.}~\bibnamefont {Travesset}},\
  }\bibfield  {title} {\bibinfo {title} {Interacting topological defects on
  frozen topographies},\ }\href {https://doi.org/10.1103/PhysRevB.62.8738}
  {\bibfield  {journal} {\bibinfo  {journal} {Phys. Rev. B}\ }\textbf {\bibinfo
  {volume} {62}},\ \bibinfo {pages} {8738} (\bibinfo {year}
  {2000})}\BibitemShut {NoStop}%
\bibitem [{\citenamefont {Giomi}\ and\ \citenamefont
  {Bowick}(2007)}]{Giomi2007}%
  \BibitemOpen
  \bibfield  {author} {\bibinfo {author} {\bibfnamefont {L.}~\bibnamefont
  {Giomi}}\ and\ \bibinfo {author} {\bibfnamefont {M.}~\bibnamefont {Bowick}},\
  }\bibfield  {title} {\bibinfo {title} {Crystalline order on {Riemannian}
  manifolds with variable {Gaussian} curvature and boundary},\ }\href
  {https://doi.org/10.1103/PhysRevB.76.054106} {\bibfield  {journal} {\bibinfo
  {journal} {Phys. Rev. B}\ }\textbf {\bibinfo {volume} {76}},\ \bibinfo
  {pages} {054106} (\bibinfo {year} {2007})}\BibitemShut {NoStop}%
\bibitem [{\citenamefont {Bowick}\ and\ \citenamefont
  {Giomi}(2009)}]{Bowick2009}%
  \BibitemOpen
  \bibfield  {author} {\bibinfo {author} {\bibfnamefont {M.~J.}\ \bibnamefont
  {Bowick}}\ and\ \bibinfo {author} {\bibfnamefont {L.}~\bibnamefont {Giomi}},\
  }\bibfield  {title} {\bibinfo {title} {Two-dimensional matter: {Order},
  curvature and defects},\ }\href {https://doi.org/10.1080/00018730903043166}
  {\bibfield  {journal} {\bibinfo  {journal} {Adv. Phys.}\ }\textbf {\bibinfo
  {volume} {58}},\ \bibinfo {pages} {449} (\bibinfo {year} {2009})}\BibitemShut
  {NoStop}%
\bibitem [{\citenamefont {Pedley}\ and\ \citenamefont
  {Kessler}(1992)}]{Pedley1992}%
  \BibitemOpen
  \bibfield  {author} {\bibinfo {author} {\bibfnamefont {T.~J.}\ \bibnamefont
  {Pedley}}\ and\ \bibinfo {author} {\bibfnamefont {J.~O.}\ \bibnamefont
  {Kessler}},\ }\bibfield  {title} {\bibinfo {title} {Hydrodynamic {Phenomena}
  in {Suspensions} of {Swimming} {Microorganisms}},\ }\href
  {https://doi.org/10.1146/annurev.fl.24.010192.001525} {\bibfield  {journal}
  {\bibinfo  {journal} {Annu. Rev. Fluid Mech.}\ }\textbf {\bibinfo {volume}
  {24}},\ \bibinfo {pages} {313} (\bibinfo {year} {1992})}\BibitemShut
  {NoStop}%
\bibitem [{\citenamefont {Aditi~Simha}\ and\ \citenamefont
  {Ramaswamy}(2002)}]{Simha2002}%
  \BibitemOpen
  \bibfield  {author} {\bibinfo {author} {\bibfnamefont {R.}~\bibnamefont
  {Aditi~Simha}}\ and\ \bibinfo {author} {\bibfnamefont {S.}~\bibnamefont
  {Ramaswamy}},\ }\bibfield  {title} {\bibinfo {title} {Hydrodynamic
  fluctuations and instabilities in ordered suspensions of self-propelled
  particles},\ }\href {https://doi.org/10.1103/PhysRevLett.89.058101}
  {\bibfield  {journal} {\bibinfo  {journal} {Phys. Rev. Lett.}\ }\textbf
  {\bibinfo {volume} {89}},\ \bibinfo {pages} {058101} (\bibinfo {year}
  {2002})}\BibitemShut {NoStop}%
\bibitem [{\citenamefont {Asano}\ \emph {et~al.}(2009)\citenamefont {Asano},
  \citenamefont {Jiménez‐Dalmaroni}, \citenamefont {Liverpool},
  \citenamefont {Marchetti}, \citenamefont {Giomi}, \citenamefont {Kiger},
  \citenamefont {Duke},\ and\ \citenamefont {Baum}}]{Asano2009}%
  \BibitemOpen
  \bibfield  {author} {\bibinfo {author} {\bibfnamefont {Y.}~\bibnamefont
  {Asano}}, \bibinfo {author} {\bibfnamefont {A.}~\bibnamefont
  {Jiménez‐Dalmaroni}}, \bibinfo {author} {\bibfnamefont {T.~B.}\
  \bibnamefont {Liverpool}}, \bibinfo {author} {\bibfnamefont {M.~C.}\
  \bibnamefont {Marchetti}}, \bibinfo {author} {\bibfnamefont {L.}~\bibnamefont
  {Giomi}}, \bibinfo {author} {\bibfnamefont {A.}~\bibnamefont {Kiger}},
  \bibinfo {author} {\bibfnamefont {T.}~\bibnamefont {Duke}},\ and\ \bibinfo
  {author} {\bibfnamefont {B.}~\bibnamefont {Baum}},\ }\bibfield  {title}
  {\bibinfo {title} {Pak3 inhibits local actin filament formation to regulate
  global cell polarity},\ }\href {https://doi.org/10.2976/1.3100548} {\bibfield
   {journal} {\bibinfo  {journal} {HFSP J.}\ }\textbf {\bibinfo {volume} {3}},\
  \bibinfo {pages} {194} (\bibinfo {year} {2009})}\BibitemShut {NoStop}%
\bibitem [{\citenamefont {You}\ \emph {et~al.}(2018)\citenamefont {You},
  \citenamefont {Pearce}, \citenamefont {Sengupta},\ and\ \citenamefont
  {Giomi}}]{You2018}%
  \BibitemOpen
  \bibfield  {author} {\bibinfo {author} {\bibfnamefont {Z.}~\bibnamefont
  {You}}, \bibinfo {author} {\bibfnamefont {D.~J.~G.}\ \bibnamefont {Pearce}},
  \bibinfo {author} {\bibfnamefont {A.}~\bibnamefont {Sengupta}},\ and\
  \bibinfo {author} {\bibfnamefont {L.}~\bibnamefont {Giomi}},\ }\bibfield
  {title} {\bibinfo {title} {Geometry and mechanics of microdomains in growing
  bacterial colonies},\ }\href {https://doi.org/10.1103/PhysRevX.8.031065}
  {\bibfield  {journal} {\bibinfo  {journal} {Phys. Rev. X}\ }\textbf {\bibinfo
  {volume} {8}},\ \bibinfo {pages} {031065} (\bibinfo {year}
  {2018})}\BibitemShut {NoStop}%
\bibitem [{\citenamefont {Capovilla}\ and\ \citenamefont
  {Guven}(2002)}]{Capovilla2002c}%
  \BibitemOpen
  \bibfield  {author} {\bibinfo {author} {\bibfnamefont {R.}~\bibnamefont
  {Capovilla}}\ and\ \bibinfo {author} {\bibfnamefont {J.}~\bibnamefont
  {Guven}},\ }\bibfield  {title} {\bibinfo {title} {Stresses in lipid
  membranes},\ }\href {https://doi.org/10.1088/0305-4470/35/30/302} {\bibfield
  {journal} {\bibinfo  {journal} {J. Phys. A Math. Gen.}\ }\textbf {\bibinfo
  {volume} {35}},\ \bibinfo {pages} {6233} (\bibinfo {year}
  {2002})}\BibitemShut {NoStop}%
\bibitem [{\citenamefont {Capovilla}\ and\ \citenamefont
  {Guven}(2004)}]{Capovilla2004}%
  \BibitemOpen
  \bibfield  {author} {\bibinfo {author} {\bibfnamefont {R.}~\bibnamefont
  {Capovilla}}\ and\ \bibinfo {author} {\bibfnamefont {J.}~\bibnamefont
  {Guven}},\ }\bibfield  {title} {\bibinfo {title} {Stress and geometry of
  lipid vesicles},\ }\href {https://doi.org/10.1088/0953-8984/16/22/018}
  {\bibfield  {journal} {\bibinfo  {journal} {J. Phys. Condens. Matter}\
  }\textbf {\bibinfo {volume} {16}},\ \bibinfo {pages} {S2187} (\bibinfo {year}
  {2004})}\BibitemShut {NoStop}%
\bibitem [{\citenamefont {Guven}(2006)}]{Guven2006}%
  \BibitemOpen
  \bibfield  {author} {\bibinfo {author} {\bibfnamefont {J.}~\bibnamefont
  {Guven}},\ }\bibfield  {title} {\bibinfo {title} {Laplace pressure as a
  surface stress in fluid vesicles},\ }\href
  {https://doi.org/10.1088/0305-4470/39/14/019} {\bibfield  {journal} {\bibinfo
   {journal} {J. Phys. A}\ }\textbf {\bibinfo {volume} {39}},\ \bibinfo {pages}
  {3771} (\bibinfo {year} {2006})}\BibitemShut {NoStop}%
\bibitem [{\citenamefont {Salbreux}\ and\ \citenamefont
  {Jülicher}(2017)}]{Salbreux2017}%
  \BibitemOpen
  \bibfield  {author} {\bibinfo {author} {\bibfnamefont {G.}~\bibnamefont
  {Salbreux}}\ and\ \bibinfo {author} {\bibfnamefont {F.}~\bibnamefont
  {Jülicher}},\ }\bibfield  {title} {\bibinfo {title} {Mechanics of active
  surfaces},\ }\href {https://doi.org/10.1103/PhysRevE.96.032404} {\bibfield
  {journal} {\bibinfo  {journal} {Phys. Rev. E}\ }\textbf {\bibinfo {volume}
  {96}},\ \bibinfo {pages} {032404} (\bibinfo {year} {2017})}\BibitemShut
  {NoStop}%
\bibitem [{\citenamefont {Landau}\ and\ \citenamefont
  {Lifshitz}(1970)}]{Landau1970}%
  \BibitemOpen
  \bibfield  {author} {\bibinfo {author} {\bibfnamefont {L.~D.}\ \bibnamefont
  {Landau}}\ and\ \bibinfo {author} {\bibfnamefont {E.~M.}\ \bibnamefont
  {Lifshitz}},\ }\href@noop {} {\emph {\bibinfo {title} {Theory of
  Elasticity}}}\ (\bibinfo  {publisher} {Pergamon Press},\ \bibinfo {year}
  {1970})\BibitemShut {NoStop}%
\bibitem [{\citenamefont {Seung}\ and\ \citenamefont
  {Nelson}(1988)}]{Seung1988}%
  \BibitemOpen
  \bibfield  {author} {\bibinfo {author} {\bibfnamefont {H.~S.}\ \bibnamefont
  {Seung}}\ and\ \bibinfo {author} {\bibfnamefont {D.~R.}\ \bibnamefont
  {Nelson}},\ }\bibfield  {title} {\bibinfo {title} {Defects in flexible
  membranes with crystalline order},\ }\href
  {https://doi.org/10.1103/PhysRevA.38.1005} {\bibfield  {journal} {\bibinfo
  {journal} {Phys. Rev. A}\ }\textbf {\bibinfo {volume} {38}},\ \bibinfo
  {pages} {1005} (\bibinfo {year} {1988})}\BibitemShut {NoStop}%
\bibitem [{\citenamefont {Carenza}\ \emph
  {et~al.}(2020{\natexlab{a}})\citenamefont {Carenza}, \citenamefont
  {Biferale},\ and\ \citenamefont {Gonnella}}]{carenza2020_bif}%
  \BibitemOpen
  \bibfield  {author} {\bibinfo {author} {\bibfnamefont {L.}~\bibnamefont
  {Carenza}}, \bibinfo {author} {\bibfnamefont {L.}~\bibnamefont {Biferale}},\
  and\ \bibinfo {author} {\bibfnamefont {G.}~\bibnamefont {Gonnella}},\
  }\bibfield  {title} {\bibinfo {title} {Cascade or not cascade? {Energy}
  transfer and elastic effects in active nematics},\ }\href
  {https://doi.org/10.1209/0295-5075/132/44003} {\bibfield  {journal} {\bibinfo
   {journal} {EPL}\ }\textbf {\bibinfo {volume} {132}},\ \bibinfo {pages}
  {44003} (\bibinfo {year} {2020}{\natexlab{a}})}\BibitemShut {NoStop}%
\bibitem [{\citenamefont {Kruse}\ \emph {et~al.}(2004)\citenamefont {Kruse},
  \citenamefont {Joanny}, \citenamefont {Jülicher}, \citenamefont {Prost},\
  and\ \citenamefont {Sekimoto}}]{Kruse2004}%
  \BibitemOpen
  \bibfield  {author} {\bibinfo {author} {\bibfnamefont {K.}~\bibnamefont
  {Kruse}}, \bibinfo {author} {\bibfnamefont {J.~F.}\ \bibnamefont {Joanny}},
  \bibinfo {author} {\bibfnamefont {F.}~\bibnamefont {Jülicher}}, \bibinfo
  {author} {\bibfnamefont {J.}~\bibnamefont {Prost}},\ and\ \bibinfo {author}
  {\bibfnamefont {K.}~\bibnamefont {Sekimoto}},\ }\bibfield  {title} {\bibinfo
  {title} {Asters, {Vortices}, and {Rotating} {Spirals} in {Active} {Gels} of
  {Polar} {Filaments}},\ }\href {https://doi.org/10.1103/PhysRevLett.92.078101}
  {\bibfield  {journal} {\bibinfo  {journal} {Phys. Rev. Lett.}\ }\textbf
  {\bibinfo {volume} {92}},\ \bibinfo {pages} {078101} (\bibinfo {year}
  {2004})}\BibitemShut {NoStop}%
\bibitem [{\citenamefont {Giomi}\ \emph {et~al.}(2014)\citenamefont {Giomi},
  \citenamefont {Bowick}, \citenamefont {Mishra}, \citenamefont {Sknepnek},\
  and\ \citenamefont {Marchetti}}]{Giomi2014}%
  \BibitemOpen
  \bibfield  {author} {\bibinfo {author} {\bibfnamefont {L.}~\bibnamefont
  {Giomi}}, \bibinfo {author} {\bibfnamefont {M.~J.}\ \bibnamefont {Bowick}},
  \bibinfo {author} {\bibfnamefont {P.}~\bibnamefont {Mishra}}, \bibinfo
  {author} {\bibfnamefont {R.}~\bibnamefont {Sknepnek}},\ and\ \bibinfo
  {author} {\bibfnamefont {M.~C.}\ \bibnamefont {Marchetti}},\ }\bibfield
  {title} {\bibinfo {title} {Defect dynamics in active nematics},\ }\href
  {https://doi.org/10.1098/rsta.2013.0365} {\bibfield  {journal} {\bibinfo
  {journal} {Philos. Trans. R. Soc. A}\ }\textbf {\bibinfo {volume} {372}},\
  \bibinfo {pages} {20130365} (\bibinfo {year} {2014})}\BibitemShut {NoStop}%
\bibitem [{\citenamefont {Hoffmann}\ \emph {et~al.}(2020)\citenamefont
  {Hoffmann}, \citenamefont {Schakenraad}, \citenamefont {Merks},\ and\
  \citenamefont {Giomi}}]{Hoffmann2020}%
  \BibitemOpen
  \bibfield  {author} {\bibinfo {author} {\bibfnamefont {L.~A.}\ \bibnamefont
  {Hoffmann}}, \bibinfo {author} {\bibfnamefont {K.}~\bibnamefont
  {Schakenraad}}, \bibinfo {author} {\bibfnamefont {R.~M.~H.}\ \bibnamefont
  {Merks}},\ and\ \bibinfo {author} {\bibfnamefont {L.}~\bibnamefont {Giomi}},\
  }\bibfield  {title} {\bibinfo {title} {Chiral stresses in nematic cell
  monolayers},\ }\href {https://doi.org/10.1039/C9SM01851D} {\bibfield
  {journal} {\bibinfo  {journal} {Soft Matter}\ }\textbf {\bibinfo {volume}
  {16}},\ \bibinfo {pages} {764} (\bibinfo {year} {2020})}\BibitemShut
  {NoStop}%
\bibitem [{\citenamefont {Frank}\ and\ \citenamefont
  {Kardar}(2008)}]{Frank2008}%
  \BibitemOpen
  \bibfield  {author} {\bibinfo {author} {\bibfnamefont {J.~R.}\ \bibnamefont
  {Frank}}\ and\ \bibinfo {author} {\bibfnamefont {M.}~\bibnamefont {Kardar}},\
  }\bibfield  {title} {\bibinfo {title} {Defects in nematic membranes can
  buckle into pseudospheres},\ }\href
  {https://doi.org/10.1103/PhysRevE.77.041705} {\bibfield  {journal} {\bibinfo
  {journal} {Phys. Rev. E}\ }\textbf {\bibinfo {volume} {77}},\ \bibinfo
  {pages} {041705} (\bibinfo {year} {2008})}\BibitemShut {NoStop}%
\bibitem [{\citenamefont {Carenza}\ \emph
  {et~al.}(2019{\natexlab{a}})\citenamefont {Carenza}, \citenamefont
  {Gonnella}, \citenamefont {Marenduzzo},\ and\ \citenamefont
  {Negro}}]{Carenza22065}%
  \BibitemOpen
  \bibfield  {author} {\bibinfo {author} {\bibfnamefont {L.~N.}\ \bibnamefont
  {Carenza}}, \bibinfo {author} {\bibfnamefont {G.}~\bibnamefont {Gonnella}},
  \bibinfo {author} {\bibfnamefont {D.}~\bibnamefont {Marenduzzo}},\ and\
  \bibinfo {author} {\bibfnamefont {G.}~\bibnamefont {Negro}},\ }\bibfield
  {title} {\bibinfo {title} {Rotation and propulsion in 3{D} active chiral
  droplets},\ }\href {https://www.pnas.org/content/116/44/22065} {\bibfield
  {journal} {\bibinfo  {journal} {Proc. Natl. Acad. Sci.}\ }\textbf {\bibinfo
  {volume} {116}},\ \bibinfo {pages} {22065} (\bibinfo {year}
  {2019}{\natexlab{a}})}\BibitemShut {NoStop}%
\bibitem [{\citenamefont {Carenza}\ \emph
  {et~al.}(2020{\natexlab{b}})\citenamefont {Carenza}, \citenamefont
  {Gonnella}, \citenamefont {Marenduzzo},\ and\ \citenamefont
  {Negro}}]{carenza2020_physA}%
  \BibitemOpen
  \bibfield  {author} {\bibinfo {author} {\bibfnamefont {L.}~\bibnamefont
  {Carenza}}, \bibinfo {author} {\bibfnamefont {G.}~\bibnamefont {Gonnella}},
  \bibinfo {author} {\bibfnamefont {D.}~\bibnamefont {Marenduzzo}},\ and\
  \bibinfo {author} {\bibfnamefont {G.}~\bibnamefont {Negro}},\ }\bibfield
  {title} {\bibinfo {title} {Chaotic and periodical dynamics of active chiral
  droplets},\ }\href
  {https://doi.org/https://doi.org/10.1016/j.physa.2020.125025} {\bibfield
  {journal} {\bibinfo  {journal} {Phys. A}\ }\textbf {\bibinfo {volume}
  {559}},\ \bibinfo {pages} {125025} (\bibinfo {year}
  {2020}{\natexlab{b}})}\BibitemShut {NoStop}%
\bibitem [{\citenamefont {Liu}\ and\ \citenamefont {Shen}(2003)}]{Liu2003}%
  \BibitemOpen
  \bibfield  {author} {\bibinfo {author} {\bibfnamefont {C.}~\bibnamefont
  {Liu}}\ and\ \bibinfo {author} {\bibfnamefont {J.}~\bibnamefont {Shen}},\
  }\bibfield  {title} {\bibinfo {title} {A phase field model for the mixture of
  two incompressible fluids and its approximation by a {Fourier}-spectral
  method},\ }\href
  {https://doi.org/https://doi.org/10.1016/S0167-2789(03)00030-7} {\bibfield
  {journal} {\bibinfo  {journal} {Phys. D}\ }\textbf {\bibinfo {volume}
  {179}},\ \bibinfo {pages} {211} (\bibinfo {year} {2003})}\BibitemShut
  {NoStop}%
\bibitem [{\citenamefont {Drazin}(1992)}]{drazin_1992}%
  \BibitemOpen
  \bibfield  {author} {\bibinfo {author} {\bibfnamefont {P.~G.}\ \bibnamefont
  {Drazin}},\ }\href {https://doi.org/10.1017/CBO9781139172455} {\emph
  {\bibinfo {title} {Nonlinear Systems}}},\ Cambridge Texts in Applied
  Mathematics\ (\bibinfo  {publisher} {Cambridge University Press},\ \bibinfo
  {year} {1992})\BibitemShut {NoStop}%
\bibitem [{\citenamefont {Alert}\ \emph {et~al.}(2020)\citenamefont {Alert},
  \citenamefont {Joanny},\ and\ \citenamefont {Casademunt}}]{Alert2020}%
  \BibitemOpen
  \bibfield  {author} {\bibinfo {author} {\bibfnamefont {R.}~\bibnamefont
  {Alert}}, \bibinfo {author} {\bibfnamefont {J.-F.}\ \bibnamefont {Joanny}},\
  and\ \bibinfo {author} {\bibfnamefont {J.}~\bibnamefont {Casademunt}},\
  }\bibfield  {title} {\bibinfo {title} {Universal scaling of active nematic
  turbulence},\ }\href {https://doi.org/10.1038/s41567-020-0854-4} {\bibfield
  {journal} {\bibinfo  {journal} {Nat. Phys.}\ }\textbf {\bibinfo {volume}
  {6}},\ \bibinfo {pages} {682–688} (\bibinfo {year} {2020})}\BibitemShut
  {NoStop}%
\bibitem [{\citenamefont {Carenza}\ \emph
  {et~al.}(2020{\natexlab{c}})\citenamefont {Carenza}, \citenamefont
  {Biferale},\ and\ \citenamefont {Gonnella}}]{carenza2020}%
  \BibitemOpen
  \bibfield  {author} {\bibinfo {author} {\bibfnamefont {L.~N.}\ \bibnamefont
  {Carenza}}, \bibinfo {author} {\bibfnamefont {L.}~\bibnamefont {Biferale}},\
  and\ \bibinfo {author} {\bibfnamefont {G.}~\bibnamefont {Gonnella}},\
  }\bibfield  {title} {\bibinfo {title} {Multiscale control of active emulsion
  dynamics},\ }\href {https://doi.org/10.1103/PhysRevFluids.5.011302}
  {\bibfield  {journal} {\bibinfo  {journal} {Phys. Rev. Fluids}\ }\textbf
  {\bibinfo {volume} {5}},\ \bibinfo {pages} {011302} (\bibinfo {year}
  {2020}{\natexlab{c}})}\BibitemShut {NoStop}%
\bibitem [{\citenamefont {Stone}\ and\ \citenamefont
  {Ajdari}(1998)}]{Stone1998}%
  \BibitemOpen
  \bibfield  {author} {\bibinfo {author} {\bibfnamefont {H.~A.}\ \bibnamefont
  {Stone}}\ and\ \bibinfo {author} {\bibfnamefont {A.}~\bibnamefont {Ajdari}},\
  }\bibfield  {title} {\bibinfo {title} {Hydrodynamics of particles embedded in
  a flat surfactant layer overlying a subphase of finite depth},\ }\href
  {https://doi.org/10.1017/S0022112098001980} {\bibfield  {journal} {\bibinfo
  {journal} {J. Fluid Mech.}\ }\textbf {\bibinfo {volume} {369}},\ \bibinfo
  {pages} {151–173} (\bibinfo {year} {1998})}\BibitemShut {NoStop}%
\bibitem [{\citenamefont {Diel}\ \emph {et~al.}(2020)\citenamefont {Diel},
  \citenamefont {Lichtman},\ and\ \citenamefont {Richardson}}]{Diel2020}%
  \BibitemOpen
  \bibfield  {author} {\bibinfo {author} {\bibfnamefont {E.~E.}\ \bibnamefont
  {Diel}}, \bibinfo {author} {\bibfnamefont {J.~W.}\ \bibnamefont {Lichtman}},\
  and\ \bibinfo {author} {\bibfnamefont {D.~S.}\ \bibnamefont {Richardson}},\
  }\bibfield  {title} {\bibinfo {title} {Tutorial: {Avoiding} and correcting
  sample-induced spherical aberration artifacts in {3D} fluorescence
  microscopy},\ }\href {https://doi.org/10.1038/s41596-020-0360-2} {\bibfield
  {journal} {\bibinfo  {journal} {Nat. Protoc.}\ }\textbf {\bibinfo {volume}
  {15}},\ \bibinfo {pages} {2773} (\bibinfo {year} {2020})}\BibitemShut
  {NoStop}%
\bibitem [{\citenamefont {Carenza}\ \emph
  {et~al.}(2019{\natexlab{b}})\citenamefont {Carenza}, \citenamefont
  {Gonnella}, \citenamefont {Lamura}, \citenamefont {Negro},\ and\
  \citenamefont {Tiribocchi}}]{carenza2019}%
  \BibitemOpen
  \bibfield  {author} {\bibinfo {author} {\bibfnamefont {L.~N.}\ \bibnamefont
  {Carenza}}, \bibinfo {author} {\bibfnamefont {G.}~\bibnamefont {Gonnella}},
  \bibinfo {author} {\bibfnamefont {A.}~\bibnamefont {Lamura}}, \bibinfo
  {author} {\bibfnamefont {G.}~\bibnamefont {Negro}},\ and\ \bibinfo {author}
  {\bibfnamefont {A.}~\bibnamefont {Tiribocchi}},\ }\bibfield  {title}
  {\bibinfo {title} {Lattice {B}oltzmann methods and active fluids},\ }\href
  {https://doi.org/10.1140/epje/i2019-11843-6} {\bibfield  {journal} {\bibinfo
  {journal} {Eur. Phys. J. E}\ }\textbf {\bibinfo {volume} {42}},\ \bibinfo
  {pages} {81} (\bibinfo {year} {2019}{\natexlab{b}})}\BibitemShut {NoStop}%
\end{thebibliography}%


\begin{thebibliography}{12}%
\makeatletter
\providecommand \@ifxundefined [1]{%
 \@ifx{#1\undefined}
}%
\providecommand \@ifnum [1]{%
 \ifnum #1\expandafter \@firstoftwo
 \else \expandafter \@secondoftwo
 \fi
}%
\providecommand \@ifx [1]{%
 \ifx #1\expandafter \@firstoftwo
 \else \expandafter \@secondoftwo
 \fi
}%
\providecommand \natexlab [1]{#1}%
\providecommand \enquote  [1]{``#1''}%
\providecommand \bibnamefont  [1]{#1}%
\providecommand \bibfnamefont [1]{#1}%
\providecommand \citenamefont [1]{#1}%
\providecommand \href@noop [0]{\@secondoftwo}%
\providecommand \href [0]{\begingroup \@sanitize@url \@href}%
\providecommand \@href[1]{\@@startlink{#1}\@@href}%
\providecommand \@@href[1]{\endgroup#1\@@endlink}%
\providecommand \@sanitize@url [0]{\catcode `\\12\catcode `\$12\catcode
  `\&12\catcode `\#12\catcode `\^12\catcode `\_12\catcode `\%12\relax}%
\providecommand \@@startlink[1]{}%
\providecommand \@@endlink[0]{}%
\providecommand \url  [0]{\begingroup\@sanitize@url \@url }%
\providecommand \@url [1]{\endgroup\@href {#1}{\urlprefix }}%
\providecommand \urlprefix  [0]{URL }%
\providecommand \Eprint [0]{\href }%
\providecommand \doibase [0]{https://doi.org/}%
\providecommand \selectlanguage [0]{\@gobble}%
\providecommand \bibinfo  [0]{\@secondoftwo}%
\providecommand \bibfield  [0]{\@secondoftwo}%
\providecommand \translation [1]{[#1]}%
\providecommand \BibitemOpen [0]{}%
\providecommand \bibitemStop [0]{}%
\providecommand \bibitemNoStop [0]{.\EOS\space}%
\providecommand \EOS [0]{\spacefactor3000\relax}%
\providecommand \BibitemShut  [1]{\csname bibitem#1\endcsname}%
\let\auto@bib@innerbib\@empty
\bibitem [{\citenamefont {Diel}\ \emph {et~al.}(2020)\citenamefont {Diel},
  \citenamefont {Lichtman},\ and\ \citenamefont {Richardson}}]{Diel2020}%
  \BibitemOpen
  \bibfield  {author} {\bibinfo {author} {\bibfnamefont {E.~E.}\ \bibnamefont
  {Diel}}, \bibinfo {author} {\bibfnamefont {J.~W.}\ \bibnamefont {Lichtman}},\
  and\ \bibinfo {author} {\bibfnamefont {D.~S.}\ \bibnamefont {Richardson}},\
  }\bibfield  {title} {\bibinfo {title} {Tutorial: {Avoiding} and correcting
  sample-induced spherical aberration artifacts in {3D} fluorescence
  microscopy},\ }\href {https://doi.org/10.1038/s41596-020-0360-2} {\bibfield
  {journal} {\bibinfo  {journal} {Nat. Protoc.}\ }\textbf {\bibinfo {volume}
  {15}},\ \bibinfo {pages} {2773} (\bibinfo {year} {2020})}\BibitemShut
  {NoStop}%
\bibitem [{\citenamefont {Latorre}\ \emph {et~al.}(2018)\citenamefont
  {Latorre}, \citenamefont {Kale}, \citenamefont {Casares}, \citenamefont
  {Gómez-González}, \citenamefont {Uroz}, \citenamefont {Valon},
  \citenamefont {Nair}, \citenamefont {Garreta}, \citenamefont {Montserrat},
  \citenamefont {del Campo}, \citenamefont {Ladoux}, \citenamefont {Arroyo},\
  and\ \citenamefont {Trepat}}]{Latorre2018}%
  \BibitemOpen
  \bibfield  {author} {\bibinfo {author} {\bibfnamefont {E.}~\bibnamefont
  {Latorre}}, \bibinfo {author} {\bibfnamefont {S.}~\bibnamefont {Kale}},
  \bibinfo {author} {\bibfnamefont {L.}~\bibnamefont {Casares}}, \bibinfo
  {author} {\bibfnamefont {M.}~\bibnamefont {Gómez-González}}, \bibinfo
  {author} {\bibfnamefont {M.}~\bibnamefont {Uroz}}, \bibinfo {author}
  {\bibfnamefont {L.}~\bibnamefont {Valon}}, \bibinfo {author} {\bibfnamefont
  {R.~V.}\ \bibnamefont {Nair}}, \bibinfo {author} {\bibfnamefont
  {E.}~\bibnamefont {Garreta}}, \bibinfo {author} {\bibfnamefont
  {N.}~\bibnamefont {Montserrat}}, \bibinfo {author} {\bibfnamefont
  {A.}~\bibnamefont {del Campo}}, \bibinfo {author} {\bibfnamefont
  {B.}~\bibnamefont {Ladoux}}, \bibinfo {author} {\bibfnamefont
  {M.}~\bibnamefont {Arroyo}},\ and\ \bibinfo {author} {\bibfnamefont
  {X.}~\bibnamefont {Trepat}},\ }\bibfield  {title} {\bibinfo {title} {Active
  superelasticity in three-dimensional epithelia of controlled shape},\ }\href
  {https://doi.org/10.1038/s41586-018-0671-4} {\bibfield  {journal} {\bibinfo
  {journal} {Nature}\ }\textbf {\bibinfo {volume} {563}},\ \bibinfo {pages}
  {203} (\bibinfo {year} {2018})}\BibitemShut {NoStop}%
\bibitem [{\citenamefont {Wyatt}\ \emph {et~al.}(2020)\citenamefont {Wyatt},
  \citenamefont {Fouchard}, \citenamefont {Lisica}, \citenamefont
  {Khalilgharibi}, \citenamefont {Baum}, \citenamefont {Recho}, \citenamefont
  {Kabla},\ and\ \citenamefont {Charras}}]{Wyatt2020}%
  \BibitemOpen
  \bibfield  {author} {\bibinfo {author} {\bibfnamefont {T.~P.~J.}\
  \bibnamefont {Wyatt}}, \bibinfo {author} {\bibfnamefont {J.}~\bibnamefont
  {Fouchard}}, \bibinfo {author} {\bibfnamefont {A.}~\bibnamefont {Lisica}},
  \bibinfo {author} {\bibfnamefont {N.}~\bibnamefont {Khalilgharibi}}, \bibinfo
  {author} {\bibfnamefont {B.}~\bibnamefont {Baum}}, \bibinfo {author}
  {\bibfnamefont {P.}~\bibnamefont {Recho}}, \bibinfo {author} {\bibfnamefont
  {A.~J.}\ \bibnamefont {Kabla}},\ and\ \bibinfo {author} {\bibfnamefont
  {G.~T.}\ \bibnamefont {Charras}},\ }\bibfield  {title} {\bibinfo {title}
  {Actomyosin controls planarity and folding of epithelia in response to
  compression},\ }\href {https://doi.org/10.1038/s41563-019-0461-x} {\bibfield
  {journal} {\bibinfo  {journal} {Nat. Mater.}\ }\textbf {\bibinfo {volume}
  {19}},\ \bibinfo {pages} {109} (\bibinfo {year} {2020})}\BibitemShut
  {NoStop}%
\bibitem [{\citenamefont {Cereijido}\ \emph {et~al.}(1978)\citenamefont
  {Cereijido}, \citenamefont {Robbins}, \citenamefont {Dolan}, \citenamefont
  {Rotunno},\ and\ \citenamefont {Sabatini}}]{Cereijido1978}%
  \BibitemOpen
  \bibfield  {author} {\bibinfo {author} {\bibfnamefont {M.}~\bibnamefont
  {Cereijido}}, \bibinfo {author} {\bibfnamefont {E.}~\bibnamefont {Robbins}},
  \bibinfo {author} {\bibfnamefont {W.}~\bibnamefont {Dolan}}, \bibinfo
  {author} {\bibfnamefont {C.}~\bibnamefont {Rotunno}},\ and\ \bibinfo {author}
  {\bibfnamefont {D.}~\bibnamefont {Sabatini}},\ }\bibfield  {title} {\bibinfo
  {title} {{Polarized monolayers formed by epithelial cells on a permeable and
  translucent support}},\ }\href {https://doi.org/10.1083/jcb.77.3.853}
  {\bibfield  {journal} {\bibinfo  {journal} {J. Cell Biol.}\ }\textbf
  {\bibinfo {volume} {77}},\ \bibinfo {pages} {853} (\bibinfo {year}
  {1978})}\BibitemShut {NoStop}%
\bibitem [{\citenamefont {Salbreux}\ and\ \citenamefont
  {Jülicher}(2017)}]{Salbreux2017}%
  \BibitemOpen
  \bibfield  {author} {\bibinfo {author} {\bibfnamefont {G.}~\bibnamefont
  {Salbreux}}\ and\ \bibinfo {author} {\bibfnamefont {F.}~\bibnamefont
  {Jülicher}},\ }\bibfield  {title} {\bibinfo {title} {Mechanics of active
  surfaces},\ }\href {https://doi.org/10.1103/PhysRevE.96.032404} {\bibfield
  {journal} {\bibinfo  {journal} {Phys. Rev. E}\ }\textbf {\bibinfo {volume}
  {96}},\ \bibinfo {pages} {032404} (\bibinfo {year} {2017})}\BibitemShut
  {NoStop}%
\bibitem [{\citenamefont {Mietke}\ \emph {et~al.}(2019)\citenamefont {Mietke},
  \citenamefont {Jemseena}, \citenamefont {Kumar}, \citenamefont {Sbalzarini},\
  and\ \citenamefont {Jülicher}}]{Mietke2019b}%
  \BibitemOpen
  \bibfield  {author} {\bibinfo {author} {\bibfnamefont {A.}~\bibnamefont
  {Mietke}}, \bibinfo {author} {\bibfnamefont {V.}~\bibnamefont {Jemseena}},
  \bibinfo {author} {\bibfnamefont {K.~V.}\ \bibnamefont {Kumar}}, \bibinfo
  {author} {\bibfnamefont {I.~F.}\ \bibnamefont {Sbalzarini}},\ and\ \bibinfo
  {author} {\bibfnamefont {F.}~\bibnamefont {Jülicher}},\ }\bibfield  {title}
  {\bibinfo {title} {Minimal {Model} of {Cellular} {Symmetry} {Breaking}},\
  }\href {https://doi.org/10.1103/PhysRevLett.123.188101} {\bibfield  {journal}
  {\bibinfo  {journal} {Phys. Rev. Lett.}\ }\textbf {\bibinfo {volume} {123}},\
  \bibinfo {pages} {188101} (\bibinfo {year} {2019})}\BibitemShut {NoStop}%
\bibitem [{\citenamefont {Capovilla}\ and\ \citenamefont
  {Guven}(2002)}]{Capovilla2002c}%
  \BibitemOpen
  \bibfield  {author} {\bibinfo {author} {\bibfnamefont {R.}~\bibnamefont
  {Capovilla}}\ and\ \bibinfo {author} {\bibfnamefont {J.}~\bibnamefont
  {Guven}},\ }\bibfield  {title} {\bibinfo {title} {Stresses in lipid
  membranes},\ }\href {https://doi.org/10.1088/0305-4470/35/30/302} {\bibfield
  {journal} {\bibinfo  {journal} {J. Phys. A Math. Gen.}\ }\textbf {\bibinfo
  {volume} {35}},\ \bibinfo {pages} {6233} (\bibinfo {year}
  {2002})}\BibitemShut {NoStop}%
\bibitem [{\citenamefont {Capovilla}\ and\ \citenamefont
  {Guven}(2004)}]{Capovilla2004}%
  \BibitemOpen
  \bibfield  {author} {\bibinfo {author} {\bibfnamefont {R.}~\bibnamefont
  {Capovilla}}\ and\ \bibinfo {author} {\bibfnamefont {J.}~\bibnamefont
  {Guven}},\ }\bibfield  {title} {\bibinfo {title} {Stress and geometry of
  lipid vesicles},\ }\href {https://doi.org/10.1088/0953-8984/16/22/018}
  {\bibfield  {journal} {\bibinfo  {journal} {J. Phys. Condens. Matter}\
  }\textbf {\bibinfo {volume} {16}},\ \bibinfo {pages} {S2187} (\bibinfo {year}
  {2004})}\BibitemShut {NoStop}%
\bibitem [{\citenamefont {Guven}(2006)}]{Guven2006}%
  \BibitemOpen
  \bibfield  {author} {\bibinfo {author} {\bibfnamefont {J.}~\bibnamefont
  {Guven}},\ }\bibfield  {title} {\bibinfo {title} {Laplace pressure as a
  surface stress in fluid vesicles},\ }\href
  {https://doi.org/10.1088/0305-4470/39/14/019} {\bibfield  {journal} {\bibinfo
   {journal} {J. Phys. A}\ }\textbf {\bibinfo {volume} {39}},\ \bibinfo {pages}
  {3771} (\bibinfo {year} {2006})}\BibitemShut {NoStop}%
\bibitem [{\citenamefont {Santiago}(2018)}]{Santiago2018}%
  \BibitemOpen
  \bibfield  {author} {\bibinfo {author} {\bibfnamefont {J.~A.}\ \bibnamefont
  {Santiago}},\ }\bibfield  {title} {\bibinfo {title} {Stresses in curved
  nematic membranes},\ }\href {https://doi.org/10.1103/PhysRevE.97.052706}
  {\bibfield  {journal} {\bibinfo  {journal} {Phys. Rev. E}\ }\textbf {\bibinfo
  {volume} {97}},\ \bibinfo {pages} {052706} (\bibinfo {year}
  {2018})}\BibitemShut {NoStop}%
\bibitem [{\citenamefont {Carenza}\ \emph {et~al.}(2020)\citenamefont
  {Carenza}, \citenamefont {Biferale},\ and\ \citenamefont
  {Gonnella}}]{carenza2020}%
  \BibitemOpen
  \bibfield  {author} {\bibinfo {author} {\bibfnamefont {L.~N.}\ \bibnamefont
  {Carenza}}, \bibinfo {author} {\bibfnamefont {L.}~\bibnamefont {Biferale}},\
  and\ \bibinfo {author} {\bibfnamefont {G.}~\bibnamefont {Gonnella}},\
  }\bibfield  {title} {\bibinfo {title} {Multiscale control of active emulsion
  dynamics},\ }\href {https://doi.org/10.1103/PhysRevFluids.5.011302}
  {\bibfield  {journal} {\bibinfo  {journal} {Phys. Rev. Fluids}\ }\textbf
  {\bibinfo {volume} {5}},\ \bibinfo {pages} {011302} (\bibinfo {year}
  {2020})}\BibitemShut {NoStop}%
\bibitem [{\citenamefont {Carenza}\ \emph {et~al.}(2019)\citenamefont
  {Carenza}, \citenamefont {Gonnella}, \citenamefont {Marenduzzo},\ and\
  \citenamefont {Negro}}]{Carenza22065}%
  \BibitemOpen
  \bibfield  {author} {\bibinfo {author} {\bibfnamefont {L.~N.}\ \bibnamefont
  {Carenza}}, \bibinfo {author} {\bibfnamefont {G.}~\bibnamefont {Gonnella}},
  \bibinfo {author} {\bibfnamefont {D.}~\bibnamefont {Marenduzzo}},\ and\
  \bibinfo {author} {\bibfnamefont {G.}~\bibnamefont {Negro}},\ }\bibfield
  {title} {\bibinfo {title} {Rotation and propulsion in 3{D} active chiral
  droplets},\ }\href {https://www.pnas.org/content/116/44/22065} {\bibfield
  {journal} {\bibinfo  {journal} {Proc. Natl. Acad. Sci.}\ }\textbf {\bibinfo
  {volume} {116}},\ \bibinfo {pages} {22065} (\bibinfo {year}
  {2019})}\BibitemShut {NoStop}%
\end{thebibliography}%

\end{document}


\title{Theory of defect-mediated morphogenesis: Supplementary information}
\author{Ludwig A. Hoffmann$^1$}
\author{Livio Nicola Carenza$^1$}
\author{Julia Eckert$^2$}
\author{Luca Giomi$^1$}
\email{giomi@lorentz.leidenuniv.nl}
\affiliation{$^1$ Instituut-Lorentz, Universiteit Leiden, P.O. Box 9506, 2300 RA Leiden, Netherlands}
\affiliation{$^2$ Physics of Life Processes, Leiden Institute of Physics, Universiteit Leiden,  P.O. Box 9506, 2300 RA Leiden, Netherlands}
\date{\today}

\maketitle

\section{Experimental details.}
\textbf{Cell culture.} Parental Madin-Darby Canine Kidney (MDCK) GII cells (kindly provided by M. Gloerich, UMC Utrecht) were cultured in a 1:1 ratio of low glucose DMEM (D6046; Sigma-Aldrich) and Nutrient Mixture F-12 Ham (N4888; Sigma-Aldrich) supplemented with 10\% fetal calf serum (Thermo Fisher Scientific), and 100 mg/mL penicillin/streptomycin, 37 $^\circ$C, 5\% CO$_2$. For experiments, cells were seeded on non-coated cover glasses and cultured in high-glucose Dulbecco Modified Eagle’s Medium (D1145; Sigma-Aldrich) supplemented with 10\% fetal calf serum, 2 mM glutamine, and 100 mg/mL penicillin/streptomycin. Before fixation, cells were incubated for 2 h in the CDK1-inhibitor RO-3306 (10 $\mu$M in final concentration; SML0569; Sigma-Aldrich).

\textbf{Immunostaining.} After 8 h, cells formed a closed monolayer. To increase the cell density within the monolayer and to attain a buckling instability, cells were cultured for a total of 25 h and then fixed for 15 min in 4\% paraformaldehyde (43368; Alfa Aesar) in phosphate-buffered saline (PBS). After fixation, cells were permeabilized for 10 min with 0.1\% Triton-X 100 and blocked for 60 min with 1\% bovine serum albumin in PBS. E-cadherin was visualized using an E-cadherin rabbit antibody (1:500 ratio; 24E10; Cell Signalling) followed by staining with Alexa 532 antirabbit secondary antibody (1:500 ratio; A-11009; Invitrogen). F-actin was stained with Alexa Fluor 647-labeled phalloidin (1:500 ratio; A22287; Invitrogen) and the DNA with DAPI (1:1000 ratio; Sigma- Aldrich). 

\textbf{Microscopy.} Samples were imaged at high resolution on a home-build optical microscope setup based on an inverted Axiovert200 microscope body (Carl Zeiss, Oberkochen, Germany), a spinning disk unit (CSU-X1; Yokogawa Electric, Musashino, Tokyo, Japan), and an emCCD camera (iXon 897; Andor Labs, Morrisville, NC). IQ-software (Andor Labs) was used for setup-control and data acquisition. Illumination was performed using fiber-coupling of different lasers (405 nm (CrystaLaser, Reno, NV), 514 nm (Cobolt AB, Solna, Sweden), and 642 nm (Spectra-Physics Excelsior; Spectra-Physics, Stahnsdorf, Germany)). Cells adhered on a cover glass were inspected with an EC Plan-NEOFLUAR 40 1.3 Oil Immersion Objective (Carl Zeiss).

\textbf{Image analysis.}
Cell segmentations and the height-to-radius profile analysis were performed using written scripts in Matlab2018a. Cell boundaries were identified from a maximum intensity projection of the F-actin signal of a confocal z-stack of the top part of the dome. The height profile was determined by the averaged intensity of the F-actin signal of a radius-dependent annulus area per plane of the z-stack. Fiji software was used for the orthogonal view of the dome. 3D reconstructions were done by ImarisViewer9.7.0 and z-directions were corrected for spherical aberration and axial distortion~\cite{Diel2020}.

\section{Height function of MDCK domes}

\begin{figure*}[t!]
\centering
{\includegraphics[width=0.8\textwidth]{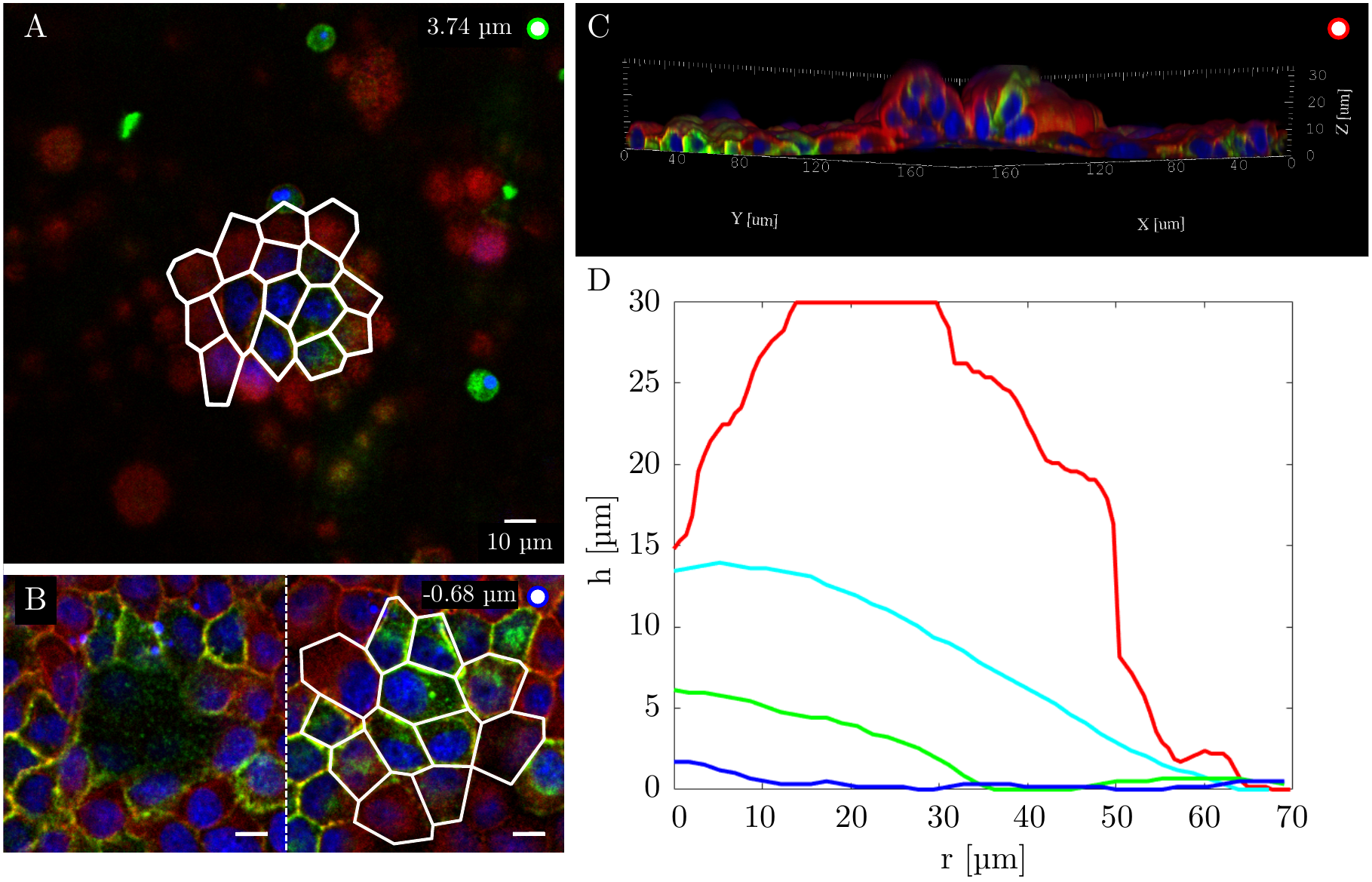}}
\caption{\textbf{Domes in MDCK layers. }Pictures of three domes of different sizes in panels (A)-(C) for which we measured the height of the cell monolayer above the flat substrate. The height of the outer area of domes as a function of the radius (distance from the center of the dome along the flat reference plane) is shown in panel (D). The color of each plot corresponds to the label in the panels (A)-(C), cyan to the dome in Fig. 1D in the main text. Segmented cells of domes shown in panel (A) and (B) illustrate the presence of topological defects near the top of the dome. For the three smallest domes, namely the functions colored blue [shown in panel (B)], green [shown in panel (A)], and cyan (shown in Fig 1D in the main text), we see that with increasing height at the center both the radius of the dome and the width of the plateau of constant height near the center are increasing. Furthermore, the height is a monotonically decreasing function of the radius. Instead, for the largest dome [red and shown in panel (D)] additional chaotic and folded structures were observed on top of the dome such that the height is maximal away from the origin. Red: F-actin, green: E-cadherin, blue: nuclei.}
\label{fig:figS1}
\end{figure*}

\begin{figure*}
\centering
{\includegraphics[width=\textwidth]{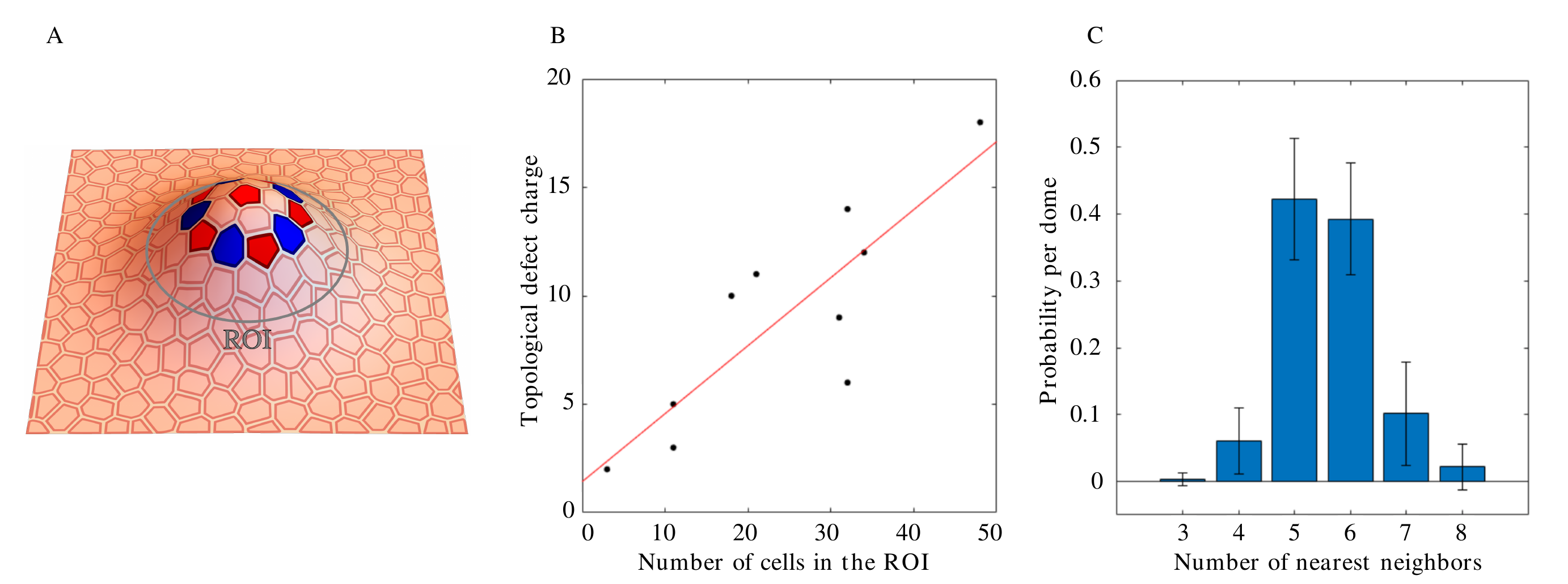}}
\caption{\textbf{Topological charge of domes. }(A) Sketch of topological defects on a dome. Heptagons are blue and pentagons are red. (B) The total topological defect charge per dome increases linearly with the number of cells constituting the dome. (C) The cells contributing a positive topological charge are mainly pentagons. The height of each bar in the histogram is the averaged probability of finding a cell with $n$ nearest neighbors in the eleven analysed domes. The error bar represents their standard deviations.}
\label{fig:figS2}
\end{figure*}

MDCK GII epithelial cells were growing on a coverslip. We observe that, after they formed a closed monolayer on the substrate, non-planar features, so called domes, developed due to the superelastic properties of the cells ~\cite{Latorre2018}. We imaged domes of different heights and diameters (see Fig.~\ref{fig:figS1}). Depending on the focal plane between the inner and outer area of the monolayer, cells changed their cell-cell contact length and the number of nearest neighbors. As described in the main text, this leads to the presence of topological defects in regions of high curvature (Fig.~\ref{fig:figS1}A and B). Furthermore, we sometimes found the additional disordered and folded structures on top of domes, see for example Fig.~\ref{fig:figS1}C. We measured the height of the buckled cell layer above the flat substrate. The height-to-radius profile of the outer area of the monolayer shows with increasing radius an increase in height, see Fig. \ref{fig:figS1}D. We assume that the buckling instability and the formation of additional structures on top of the dome is caused by the cells continued division and growth even after they form a closed monolayer. The growth and division would cause the buildup of stresses in the monolayer that are relieved by the buckling. This is in agreement with e.g. Ref. \cite{Wyatt2020}. We speculate that, consistently with the mechanism outlined in the main text, a net positive topological charge could facilitate the out-of-plane deformation of the monolayer, hence the formation of domes, in combination with other system-specific mechanisms, such as the injection of fluid under the cell layer which results in a focal detachment \cite{Cereijido1978,Latorre2018}. Furthermore, it is possible that a difference in ion-concentration surrounding the monolayer between apical and basal side can even support the dome formation in areas where topological defects appear, see also Ref. \cite{Cereijido1978}.

To support this speculation, we have we analyzed eleven domes and counted the total topological charge in the central region, where the Gaussian curvature is maximal and positive. The topological charge of a cell having $c_{i}$ sides is conventionally defined as $q_{i}=6-c_{i}$. Thus, pentagonal cells (i.e. $c_{i}=5$) have topological charge $q_{i}=1$, whereas heptagonal cells (i.e. $c_{i}=7$) has charge $q_{i}=-1$. The total topological charge is then computed by summing the individual charges of all the cells in the central region of a dome, hereafter referred to as ``region of interest'' (ROI): $Q=\sum_{i\in{\rm ROI}}q_{i}$ (see Fig.~\ref{fig:figS2}A). We found that all domes in our sample feature a positive total topological defect charge, with the mean charge of the eleven domes being $Q_\text{mean}=9.3 \pm 4.9$ (mean $\pm$ standard deviation).

Furthermore, as shown in  Fig.~\ref{fig:figS2}B we find that the total topological charge of a dome is strongly correlated with the total number of cells per dome (linear correlation coefficient $r=0.85$). Such a linear relation, is consistent with the hypothesis that the topological charge of the cellular monolayer effectively screened by its Gaussian curvature, i.e.
\begin{equation}\label{eq:screening}
Q \approx \int_{\rm dome} {\rm d}A\,K\;.	
\end{equation}
To illustrate this point, one can approximate the dome as a spherical cap of radius $R$, so that $K=1/R^{2}$ and ${\rm d}A={\rm d}\Omega\,R^{2}$, with ${\rm d}\Omega$ the infinitesimal solid angle. The right hand side of Eq. \eqref{eq:screening} equates then the solid angle $\Delta\Omega$ spanned by the ROI: i.e. $Q \approx \Delta\Omega$ (see Fig.~\ref{fig:figS2}A). This, in turn, is proportional to the number of cells it encloses, given that $N_{\rm cells} = A_{\rm ROI}/A_{\rm cell} = R^{2}\Delta\Omega/A_{\rm cell}$. Thus
\begin{equation}
Q \approx \left(\frac{A_{\rm cell}}{R^{2}}\right)N_{\rm cells}\;.	
\end{equation}
Fig.~\ref{fig:figS2}C shows instead the probability distribution of the number of sides $c_{i}$ of the cells in the ROI obtained from a sample of eleven domes. We see that a large majority of the cells are either pentagonal or hexagonal with almost half of the observed population being pentagonal (probability of $0.42\pm0.09$ (mean $\pm$ standard deviation)). Since, defects larger than $6-$fold cells are underrepresented this leads to all the domes observed having a positive total topological charge.

\section{Force Balance}

In general, the force balance on the surface is given by $\nabla_i \bm{\sigma}^i = - \bm{\Xi}^\text{ext}$ \cite{Salbreux2017, Mietke2019b},where $\bm{\sigma}^i$ denotes the surface stress and $\bm{\Xi}^\text{ext}$ external forces applied to the system. We use bold letters to refer to vectors in $\mathbb{R}^3$ and latin indices to refer to surface coordinates on $\mathcal{M}$. For example, the stress tensor can be written as a $2 \times 3$ matrix. It is possible to decompose the stress into its tangential and normal component, $\bm{\sigma}^i = \sigma^{ij} \bm{e}_j + \sigma^i_\text{n} \bm{n}$. In the absence of external forces, the force balance can then be written as 
\begin{equation}
\nabla_i \bm \sigma^i = \left(\nabla_i \sigma^{ij} + K^{\; j}_i \sigma_{n}^i\right) \bm e_j + \left(\nabla_i \sigma_{n}^i -\sigma^{ij} K_{ij} \right) \bm n = 0\;.
\end{equation}
Here, we used $\nabla_i \bm{n} = K_i^{\; j} \bm{e}_j$ and $\nabla_j \bm{e}_j = -K_{ij} \bm{n}$. From the free energy of the main text we find, following Ref. \cite{Capovilla2002c, Capovilla2004, Guven2006, Santiago2018} the tangential and normal stress tensors
\begin{equation}
\sigma^{F}_{ij} = \gamma g_{ij} - \kappa_B H K_{ij} + \kappa_B H^2 g_{ij} +\kappa_F\left[g_{ij} \chi \nabla^2 \chi  -  \chi \nabla_i \nabla_j \chi + \nabla_i\chi\nabla_j \chi\right] \;,
\end{equation}
and
\begin{equation}
\sigma^{F}_{n,i} = \kappa_B \nabla_{i} H + \kappa_F \left[(2H g_i^{\;j} - 2K_i^{\;j})\chi + 2(K_i^j - 2 g_i^{\;j} H)\right] \nabla_j \chi + \kappa_F K_i^{\;j} \chi \nabla_j \chi \;.
\end{equation}
To derive the equations in the main text we use the well-known geometric identity $K g_{ij} = 2 H K_{ij} - K_{ik}K_j^{\; k}$ and the commutator $[\nabla_i, \nabla^2] \chi = - K \nabla_i \chi$. Using these together with the other stress tensors presented in the main text, we find the hydrodynamic equation for the momentum density and the normal force balance by projecting the general force balance onto the tangential and normal direction, respectively. We will present a more detailed derivation in a future work.
 \section{Derivation of Height Equation}
In this section we derive the height equation, Eq. (5) of the main text, from the general equations describing the dynamics of the active surface, Eqs. (2) and (3) of the main text. We work in the Monge gauge where a height field $h(r)$ is defined as described in the main text. A $+1$ defect is at the center of a disk of radius $R$ and we assume that $|\nabla h| \ll 1$. As shown in the main text, we can find the explicit solutions for the velocity and director fields given in Eqs. (4) from the Eqs. (2) of the main text and we are thus now concerned with rewriting Eq. (3) of the main text. In the small-gradient approximation we have $K \approx 0$, $H \approx -\nabla^2 h/2$, and the metric is just the identity such that covariant derivatives are equal to the flat derivatives. It is then straightforward to see that
\begin{equation}
\label{eq:ForceMembrane}
\fnm = - \gamma \nabla^2 h(r) + \frac{\kappa_B}{2} \nabla^4 h(r) \;.
\end{equation}
Furthermore, using $\chi(r) = - \log r/R$, we find
\begin{equation}
\label{eq:ForceNematic}
\fnn = \kappa_F \frac{r \partial^2_r h(r) - \partial_r h(r)}{r^3} \;.
\end{equation}
Lastly, for the curvature tensor coupled to the hydrodynamic and active stress tensor, we find the following. We have,
\begin{equation}
\label{eq:CurvatureNEStress_1}
K_{ij} \sigma^{\text{h},ij} = \left(\nabla_i\nabla_j h\right) \left[P_\text{h} \delta^{ij} - \eta \left(\nabla^i v^j + \nabla^j v^i\right) \right] \;,
\end{equation}
and
\begin{equation}
\label{eq:CurvatureNEStress_2}
K_{ij} \sigma^{\text{a},ij} = \alpha (\nabla_i\nabla_j h)  \left(\frac{1}{2} \delta^{ij} - p^i p^j \right) \;,
\end{equation}
where $\delta_{ij}$ is the flat polar metric. For the first term on the right-hand side of both Eq. (\ref{eq:CurvatureNEStress_1}) and Eq. (\ref{eq:CurvatureNEStress_2}) we have $(\nabla^i\nabla^j h) \delta_{ij} = \nabla^2 h$. Furthermore, in polar coordinates $\nabla_i\nabla_j h = \partial_i \partial_j h - \Gamma_{ij}^k \partial_k h$, with $\Gamma^k_{ij}$ the Christoffel symbol associated with $\delta_{ij}$. This is non-zero only if $i = j = r$, in which case $\nabla_r^2 h = \partial^2_r h$, or if $i = j = \varphi$, then $\nabla_\varphi^2 h = r \partial_r h$. Similarly, we have $\nabla^i v^j = \partial^i v^j + \delta^{il} \Gamma_{kl}^j v^k$. With $v_r = 0$ and $v_\varphi = v_\varphi(r)$ we then find that this is non-trivial only if $i = r$ and $j = \varphi$ such that $\nabla^r v^\varphi = \partial^r v^\varphi$. Thus, since $\nabla_r\nabla_\varphi h = 0$, the second term in Eq. (\ref{eq:CurvatureNEStress_1}) vanishes. Hence, as mentioned in the main text, the velocity does not enter explicitly in the final equation for $h$. For the last term we have:
\begin{equation}
p^i p^j\nabla_i\nabla_j h(r) = \cos^2 \epsilon \, \partial_r^2 h(r) + \frac{\sin^2\epsilon}{r} \partial_r h(r) \;,
\end{equation}
where $\epsilon = \pm\arccos(-1/\lambda)/2$ as found above.
Therefore, adding Eqs. (\ref{eq:ForceMembrane})-(\ref{eq:CurvatureNEStress_2}) together, we find
\begin{align}
0 = \fnm + \fnn + K_{ij} \left(\sigma^{\text{h},ij} + \sigma^{\text{a},ij} \right) = \left(P_\text{h}+\frac{\alpha}{2} - \gamma \right) \nabla^2 h +  \frac{\kappa_F}{r} \partial_r \frac{h}{r} + \frac{\lambda - 1}{2 \lambda} \, \partial_r^2 h + \frac{\lambda + 1}{2 \lambda r} \partial_r h + \frac{\kappa_B}{2} \nabla^4 h
\end{align}
which can be rewritten to yield the height equation, Eq. (5) in the main text.

\section{Mapping to physical units}
The model parameters in lattice units used for simulations are $ k_\phi=0.008, A_0 =0.02, \kappa_F=0.02, \beta=0.03, M=0.1, \Gamma = 1, \lambda = 1.1$, and $\eta = 5/3$.
By following previous studies~\cite{carenza2020,Carenza22065}, an approximate relation between simulation and physical units (for an active gel of cytoskeletal extracts) can be obtained using $L=1$ $\mu$m as the length-scale, $\tau=10$ ms as the time-scale, and $F=1000$ nN as the force-scale. A mapping of some relevant quantities is reported in Table~\ref{tabel_units}.
\begin{table}[ht]
\centering
{\begin{tabular}{@{}|l|l|l|@{}}
\hline
Model parameters   & Simulation units & Physical units   \\\hline
Shear viscosity, $\eta$           & $5/3$  & $1.67 \ \text{KPas}$       \\
Elastic constant, $\kappa_F$      & $0.02$  & $100 \ \text{nN}$      \\
Flow-alignment parameter, $\lambda$               & $1.1$  & dimensionless       \\
Diffusion constant, $D=Ma$        & $0.001 - 0.015$  & $0.0087 - 0.0128 \ \text{$\mu$} \text{m}^2 \text{s}^{-1}$     \\
Activity, $\alpha$    & $0-0.01$  & $(0-100) \ \text{KPa}$       \\ 
\hline
\end{tabular}\caption{Mapping of some relevant quantities between simulation units and physical units.}\label{tabel_units}}
\end{table}

\section{Movies}
\textbf{Movie 1: 3D dome animation.} 3D rotation and reconstruction of the same MDCK GII monolayer  as in Fig. 1D of the main text. Double scaling of the z-direction was chosen to ease the visualization of the buckling instability. The color code is as follows. Red: F-actin, green: E-cadherin, blue: nuclei.

\textbf{Movie 2: Profile and tessellation.} Left: top-to-bottom confocal z-stack of the same MDCK GII monolayer as in Fig.1D of the main text. At a distance of $15.3$ $\mu$m to the coverslip, cells are segmented and correlated according to their number of nearest neighbors. Top-right: cross section of the dome. The white lines indicate the inner, middle and outer area of the monolayer. Bottom-right: height-to-radius profile of the buckled monolayer starting from the center of the dome (see also Fig.~\ref{fig:figS1}). The position (red circle) follows a sigmoidal fit of the outer area and moves according to the z-stack animation in the left panel. Red: F-actin, green: E-cadherin, blue: nuclei.

\textbf{Movie 3: Oscillation between buckled states.} Oscillations between a cuspidal configuration with negative curvature at the center and a smooth configuration. 
From an initially flat state the interface buckles. Note that while the initial protrusion is growing the defect is slightly off-center, leading to the whole protrusion moving on the $xy$-plane. (The $z$-direction corresponds to the interface normal in the initial state.) After some time ($t>380000$) the interface starts to periodically oscillate between these two  configurations. The oscillation is accompanied by the growth and shrinking of additional thin protrusions at the top. The white vectors denote the polarization field, while the color code refers to the local magnitude of the flow according to the color bar on the right-hand side.

\textbf{Movie 4: Droplet nucleation.} The large extensile active stresses lead to the rapid growth of protrusions that cannot be counteracted by the elasticity of the interface. This results in a periodic breaking of the interface and the consequent nucleation of a droplet after each rupture. There are two $+1$ defects in the polarization field on each of the droplets. These deform under the straining action of the active liquid crystal on the surface and eventually dissolve due to Ostwald ripening. The white vectors denote the polarization field, while the color code refers to the local magnitude of the flow according to the color bar on the right-hand side.

\textbf{Movie 5: Chaotic dynamics.} Shown is an example of the chaotic dynamics found at very high activity. The complex dynamics in this regime are characterized by chaotic deformations of the membrane with the consequent proliferation of many protrusions at the membrane. The active stresses cannot overcome the elastic forces of the interface and the protrusions break off from the membrane and elongate under the straining effect of bending deformations in the polarization pattern. The white vectors denote the polarization field, while the color code refers to the local magnitude of the flow according to the color bar on the left-hand side.

\bibliography{Biblio.bib}